\documentclass[preprint2,longabstract,eqsecnum]{aastex}

\usepackage{bm}

\usepackage[usenames]{color}
\usepackage[breaklinks, backref, colorlinks=true]{hyperref}
\usepackage[TABBOTCAP]{subfigure}
\usepackage{epstopdf}

\usepackage{graphicx}
\graphicspath{{fig/}}
\usepackage{stmaryrd}

\def\arcm{\ifmmode {^{\scriptscriptstyle\prime}}
          \else $^{\scriptscriptstyle\prime}$\fi}
\def\pdeg{\ifmmode $\setbox0=\hbox{$^{\circ}$}\rlap{\hskip.11\wd0 .}$^{\circ}
          \else \setbox0=\hbox{$^{\circ}$}\rlap{\hskip.11\wd0 .}$^{\circ}$\fi}
\def \npix{N_{\rm pix}}
\def \nside{n_{\rm side}}
\def \nbeam{N_{\rm beam}}
\def \nlist{N_{\rm list}}
\def \nhits{N_{\rm obs}}
\def \ntod {N_{\rm tod}}

\def \urad{\hat{\mathbf{r}}}
\def \utheta{\hat{\bm{\theta}}}
\def \uphi{\hat{\bm{\phi}}}

\def \upoint{\hat{\mathbf{p}}}

\def \icap{\hat{\mathbf{i}}}
\def \jcap{\hat{\mathbf{j}}}
\def \kcap{\hat{\mathbf{k}}}

\def \ncap{\hat{\mathbf{n}}}

\def \DelOm{\Delta\Omega}

\def \d{\mathrm{d}}

\def \effBeam {\mathcal{B}}
\def \effBeamPol {\bm{\mathcal{B}}}

\def \cobs {\widetilde{C}}

\shorttitle{Fast Pixel Space Convolution: {\tt FEBeCoP}}
\shortauthors{Mitra et al.}

\begin{document}

\title{Fast Pixel Space Convolution for CMB Surveys with\\
Asymmetric Beams and Complex Scan Strategies: {\tt FEBeCoP}}

\author{S.~Mitra}
\email{Sanjit.Mitra@jpl.nasa.gov}
\affil{Jet Propulsion Laboratory, California Institute of Technology, Pasadena, CA 91109, USA}

\author{G.~Rocha}
\email{graca@caltech.edu}
\affil{Jet Propulsion Laboratory, California Institute of Technology, Pasadena, CA 91109, USA}

\author{K.~M.~G\'orski\altaffilmark{1}}
\email{Krzysztof.M.Gorski@jpl.nasa.gov}
\affil{Jet Propulsion Laboratory, California Institute of Technology, Pasadena, CA 91109, USA}

\author{K.~M.~Huffenberger\altaffilmark{2}}
\email{huffenbe@physics.miami.edu}
\affil{Department of Physics, University of Miami, 1320 Campo Sano Drive, Coral Gables, FL 33146, USA}

\author{H.~K.~Eriksen\altaffilmark{3}}
\email{h.k.k.eriksen@astro.uio.no}
\affil{Institute of Theoretical Astrophysics, University of Oslo, P.O.\ Box 1029 Blindern, N-0315 Oslo, Norway}

\author{M.~A.~J.~Ashdown\altaffilmark{4}}
\email{maja1@mrao.cam.ac.uk}
\affil{Astrophysics Group, Cavendish Laboratory, J.~J.~Thomson Avenue, Cambridge CB3 0HE, UK}

\author{C.~R.~Lawrence}
\email{Charles.R.Lawrence@jpl.nasa.gov}
\affil{Jet Propulsion Laboratory, California Institute of Technology, Pasadena, CA 91109, USA\newline}

\altaffiltext{1}{Warsaw University Observatory, Aleje Ujazdowskie 4, 00-478 Warszawa, Poland}
\altaffiltext{2}{Jet Propulsion Laboratory, California Institute of Technology, Pasadena, CA 91109, USA}
\altaffiltext{3}{Centre of Mathematics for Applications, University of Oslo, P.O.\ Box 1053, Blindern, N-0316 Oslo, Norway}
\altaffiltext{4}{Institute of Astronomy, Madingley Road, Cambridge CB3 0HA, UK}


\begin{abstract}
Precise measurement of the angular power spectrum of the Cosmic Microwave  
Background (CMB) temperature and polarization anisotropy can  
tightly constrain many cosmological models and parameters. However,  
accurate measurements can only be realized in practice provided all  
major systematic effects have been taken into account. Beam asymmetry,  
coupled with the scan strategy, is a major source of systematic error  
in scanning CMB experiments such as Planck, the focus of our current interest. 
We envision Monte Carlo  
methods to rigorously study and account for the systematic effect of  
beams in CMB analysis. Toward that goal, we have developed a fast pixel space convolution  
method that can simulate sky maps observed by a scanning instrument,  
taking into account real beam shapes and scan strategy. The essence is  
to pre-compute the ``effective beams'' using a computer code, 
``Fast Effective Beam Convolution in Pixel space'' ({\tt FEBeCoP}), that we  
have developed for the Planck mission. The code  
computes effective beams given the focal plane  
beam characteristics of the Planck instrument and the full history of  
actual satellite pointing, and performs very fast convolution of sky  
signals using the effective beams. In this paper, we describe the  
algorithm and the computational scheme that has been implemented.
We also outline a few applications of the  
effective beams in the precision analysis of Planck data, for characterizing the
CMB anisotropy and for detecting and measuring properties of point sources.
\end{abstract}


\section{Introduction}

Cosmic Microwave Background (CMB) anisotropy measurements have played a leading role in precision cosmology. Detection of the CMB monopole by Penzias and Wilson~\citep{PenziasWilson} firmly established that the Big Bang model is consistent with observation. The Cosmic Background Explorer (COBE)~\citep{COBEDMR} detected the tiny $\sim 10^{-5}$ fluctuations, which justified how structures could form in a statistically isotropic universe starting from seed density perturbations. The Wilkinson Microwave Anisotropy Probe (WMAP)~\citep{WMAP} measured cosmological parameters---e.g., the flatness of the universe---with unprecedented accuracy. There have been many suborbital experiments~\citep{toco,mill99,dasi,cbi, acbar, boom, maxima} that have probed specific aspects of CMB anisotropy. Yet only a small fraction of the information embedded in the CMB temperature anisotropy has so far been measured, and measurements of the CMB polarization anisotropy have just begun. Precise measurement of the CMB temperature and polarization anisotropy is the most important next step in building a rigorous model of the content and evolution of the universe. The Planck mission is a substantial step forward in CMB anisotropy measurement---it aims to measure almost all the information available in temperature anisotropy and will make a full-sky, high-resolution, low-noise measurement of CMB polarization anisotropy. It will also image thousands of point sources of astrophysical and cosmological interest.

Planck~\citep{bluebook06,planckmission09} is a full-sky scanning experiment launched in May 2009, taking data since August 2009. It will produce high-resolution maps with tens of millions of pixels. The Low Frequency Instrument (LFI) covers 30, 44, and 70\,GHz; the High Frequency Instrument (HFI) covers 100, 143, 216, 353, 545, and 857\,GHz.  From the second Lagrangian point of the Earth-Sun system ($L_{2}$), Planck scans nearly great circles on the sky, covering the full sky twice in one year  (\citep{dupac05}).  The satellite spins at 1\,rpm around an axis that is re-pointed roughly 30~times per day along a cycloidal path, with the spin axis moving in a 7\pdeg5 circle around the anti-Sun direction with a period of six months.  This ensures that all feeds completely cover the sky, including the ecliptic polar regions.

High-resolution, low-noise measurement of the CMB anisotropy is challenging. Cosmic-variance-limited measurement at high resolution demands not only precision, but very high accuracy---all the systematic effects must be accounted for; otherwise systematic errors can become significant or dominant compared to statistical errors. The effect of asymmetric beams is one of the most important sources of systematic error in current and future CMB missions, which invest a significant effort in measuring the polarization. 
Problems with asymmetric beam systematics,  which could transfer power from large scales to small, could be acute for  measurements of the rapidly falling power spectra at high resolution.
For polarization~\citep{Trieste08}, where the signal is an order of magnitude below the temperature anisotropy and the inherent directionality of the polarization field gets coupled to the asymmetry of the beam, the effect could lead to large deviations in the polarization anisotropy. 
In general, precision measurement of the polarization power spectrum at high-resolution requires a correction for asymmetric beams. 

\begin{figure}
\centering
\includegraphics[width=0.475\textwidth]{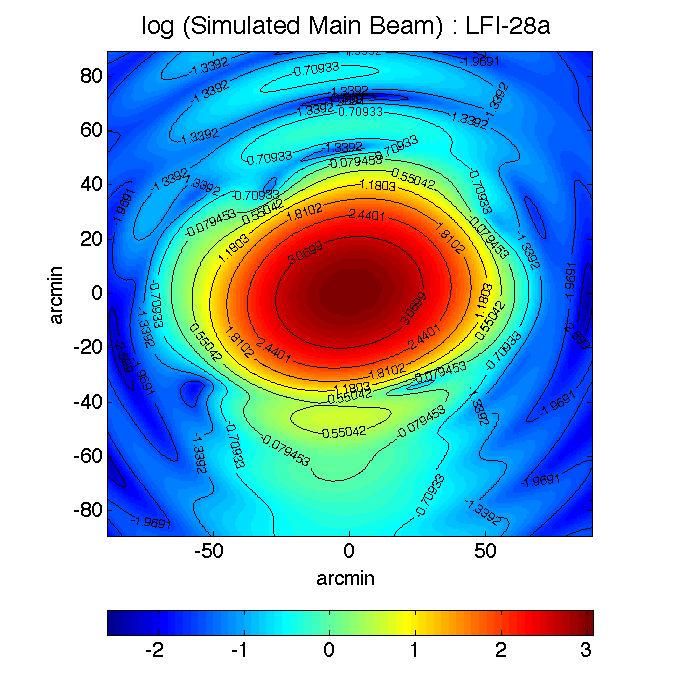}
\includegraphics[width=0.475\textwidth]{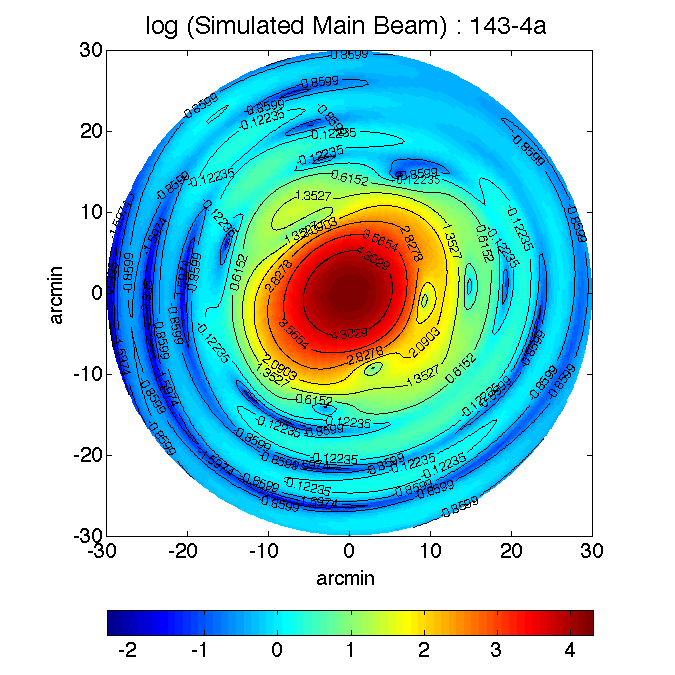}
\caption{Logarithmic plots of Planck beams for 30\,GHz (top) and 143\,GHz (bottom) as simulated by GRASP9. The 30\,GHz beam was simulated in Cartesian coordinate and the 143\,GHz one in polar coordinates. (The difference in coordinate system is just a matter of representation, it is not related to the detectors or beams.) }
\label{fig:grasp9-beams}
\end{figure}

\begin{figure}
\centering
\includegraphics[width=0.475\textwidth]{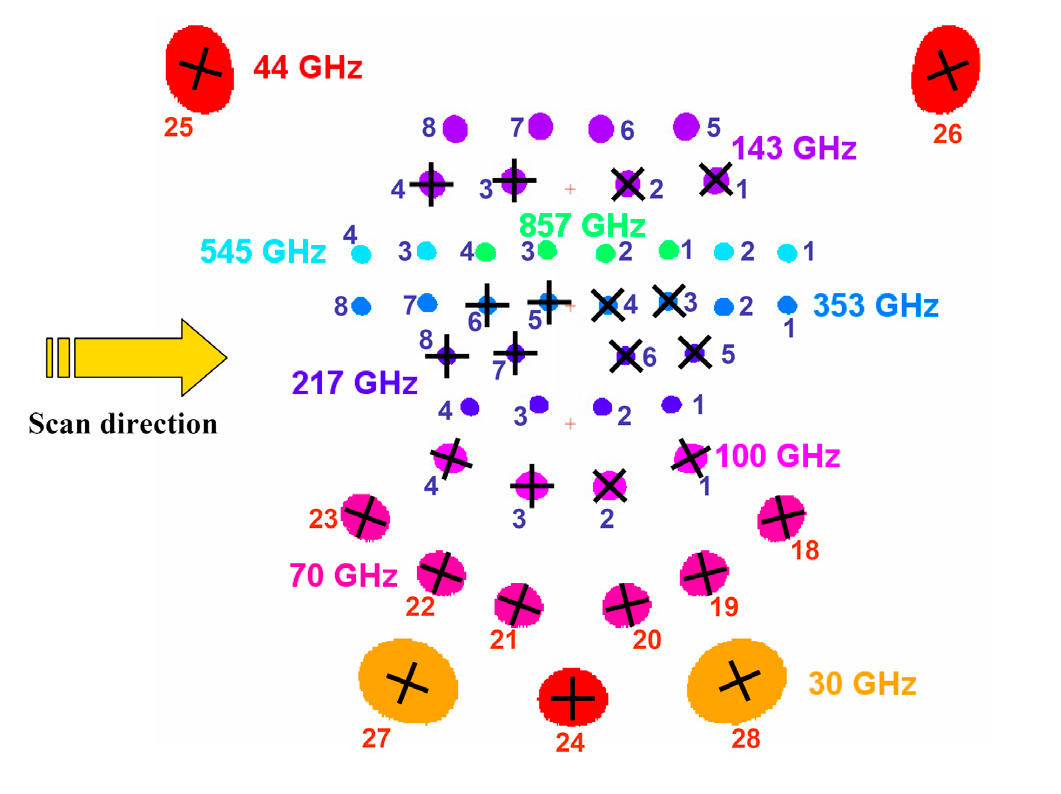}
\caption{Focal plane layout for Planck \citep{dupac05, PlanckOptics}, showing the position, approximate shape, and orientation of the beams of every detector.  Crosses show the orientation of the polarization axes of the polarization sensitive detector pairs. The arrow on the left shows the scan direction, which is fixed for the duration of the mission.}
\label{fig:focal_plane}
\end{figure}

Planck beams range from $\sim 30$\arcm at $30$ GHz to $\sim 5$\arcm at $217$GHz and higher frequencies.  The lowest frequency channels are farthest from the center of the focal plane and exhibit the most asymmetric beams.  For example, the 30\,GHz channel has  beams with ellipticities of the order 1.37.  In Figure~\ref{fig:grasp9-beams} we show realistic Planck beams, for one 30\,GHz and one 143\,GHz detector, simulated by a full diffraction analysis of the telescope using GRASP9~\citep{sandri:2002,sandri:2009, Maffei10,yurchenko:2004},
while in Figure~\ref{fig:focal_plane} we show the focal plane layout \citep{dupac05}.
The related problems of measurement of beam shapes~\citep{Huffenberger10} and the incorporation of their uncertainties in the analysis~\citep{Rocha09} are also active areas of research.  Our focus in this paper is to study the effect of asymmetries in the beam shapes on the sky maps.

Broadly speaking, the problem of asymmetric beams at high resolution has been addressed in the literature in two ways: estimating the bias of the ``pseudo-$C_l$'' estimator semi-analytically in the spherical harmonic basis~\citep{mitra09,wmap3T,mitra04,Fosalba01,TR2001}; or deconvolution of observed maps~\citep{PReBeaM, burigana03}. However, to handle a complicated instrument like Planck, with non-trivial scan strategy and noise characteristics, we aim to use a simple forward approach, i.e., to estimate the systematic effect by studying Monte-Carlo-simulated maps convolved with detector beams using the real scan path.  This requires a fast convolution algorithm for scanning CMB experiments like Planck.  Existing simulation methods~\citep{levelS}, involving spherical harmonic based convolution algorithms, are relatively slow, and even computationally prohibitive for a large number of simulations (typically a few thousand in Monte Carlo studies). This is because they generate time ordered data (TOD) prior to mapmaking.  Improving simulation capabilities was the main motivation for developing a fast, easy-to-use map convolution method which would include the effect of asymmetric beams and scan.

Convolution in pixel space is fast and flexible because in scanning CMB experiments the main beam is highly localized in pixel space.  The standard choice of pixel number $\npix$ for a given beam size ensures that only $\nbeam \sim 200$ pixels are required to accurately describe the beam. This is much smaller than the number of spherical harmonic moments needed to describe the beam with adequate accuracy. The total computation cost for pixel space convolution (after effective beam computation) scales as $\npix \, \nbeam$. This is very small compared to any other existing method. For example, using convolution in the harmonic basis for an idealized scan (where each direction has a fixed beam orientation angle), performed by the fast multipole based method ``total convolution''  \citep{WandeltGorski-00,ChallinorEtAl-00}, scales as $\sim \ell^3_\textrm{max} \, m_\textrm{max}$, where $\ell_{\rm max}$ is the highest multipole moment that contains non-negligible information about the anisotropies in the map and $m_{\rm max}$ is the highest azimuthal multipole needed to adequately describe beam asymmetry. This is more demanding than the pixel space convolution. There has been recent progress in making this step faster using recursive algebra~\citep{conviqt}. In practice, though, convolution is much more complicated than the ideal case, as the scan can take the beam to any arbitrary orientation and one has to go through the whole scan path and interpolate between orientation angles obtained from the fixed Total Convolution grid, which is a tedious process. Considering all these issues, pixel space is often a more convenient choice.

We have developed a HEALPix\footnote{\url{http://www.healpix.jpl.nasa.gov/}} \citep{HEALPix} based pixel space convolution method by pre-computing the ``effective beams'' for each pixel for both  temperature and polarization maps, incorporating asymmetric beams of the instruments and the real scan strategy. The effective beam for a given pixel is essentially the average of (discretized) beam functions for every pointing sample that fall in that pixel (a formal mathematical definition is provided later in the paper). The method has been successfully implemented for the Planck mission as a computer code {\tt FEBeCoP}, acronym of ``Fast Effective Beam Convolution in Pixel space''. {\tt FEBeCoP} efficiently utilizes large distributed computing facilities, hence it is fast, but provides a very easy interface, which enables external applications to use this facility in a convenient way. Different analyses, which require realistic Monte Carlo simulations of observed sky maps, but not TODs, can benefit from this method.

In scanning experiments which target static objects (not transient sources), the data products are better understood by studying the effective beam rather than the detector intrinsic beam function. For instance, consider an elliptic beam function that moves on the sky in a scanning pattern similar to that of Planck. If the scan is along the minor axis of the ellipse, repeated hits in a pixel will tend to make the effective beam more symmetric than the true beam, hence the final map will have less severe high resolution systematics. On the other hand, if the scan is along the major axis, the elongation will be enhanced in the effective beam and the final map will have more high resolution systematics. Effective beams thus provide a more intuitive picture of the systematic effects in the observed map as compared to the detector intrinsic beam function. This feature would be useful not only to understand the non-trivial effects of beams on observed CMB anisotropy, but also for precise measurement and extraction of point sources and Sunyaev-Zel'dovich (SZ) clusters from Planck data~\citep{bluebook06} using the accurate knowledge of the point spread function (PSF) obtained from the effective beams.

The paper is organized as follows. In section~\ref{sec:EB} we introduce effective beams. Corresponding mathematics for temperature-only and polarized effective beams are given in section~\ref{sec:math}. Section~\ref{sec:TC} analyzes the accuracy of the method comparing with theoretically exact results. Section~\ref{sec:Algo} describes the details of computational implementation, which is followed by two types of applications of the effective beams. Simulation of maps for Planck and comparison with existing simulations are shown in section~\ref{sec:conv}. Study of collective effect of the beams via elliptical Gaussian fits and estimation of approximate scalar transfer function for pseudo-$C_l$ estimator by direct spherical harmonic decomposition of beams are covered in section~\ref{sec:effBl}. We conclude with summary and discussions in section~\ref{sec:concl}.

\section{Effective Beam: Method \& Advantages}
\label{sec:EB}

The finite angular response of a detector is described by the beam function, which is closely related to the point spread function (PSF) in image space. {\em The effective beam is the average of all beam functions pointing at a certain direction for a given scan strategy}. Since, in practice, a scan strategy may not pass through exactly the same point twice, the effective beam concept makes sense only when the sky is discretized, then one can define an effective beam for each pixel on the sky.

The mapmaking algorithms~\citep{mapMakerComp,wmap3T} commonly used in CMB analyses do not take into account the finite size of the instrumental  beams, so the resulting map carries the signature of the instrumental beam function and the satellite scan strategy. In fact, {\em the observed map is a convolution of the true CMB sky with the effective beams}. Effective beams can therefore be treated as the primary object for studying the combined systematic effects of beams and scan strategy. To appreciate this fact consider the hypothetical case of symmetric detector intrinsic beam. Since the beam center (pointing direction) need not coincide with the pixel center, if the distribution of pointing directions in a pixel is elongated (say only one set of scan rings passes though that pixel), effective beams can become asymmetric.

If the effective beams were symmetric, high resolution CMB data analysis would be computationally trivial. However, in practice such an advantage is absent. The effective beams in CMB experiments are (significantly) asymmetric due to several unavoidable reasons, which include:
\begin{itemize}
\item intrinsic optics of the instruments
\item not all detectors can be placed close to the principal optic axis, so there are aberrations
\item a sample is an integral along the scan path
\item pointing directions are not uniformly distributed in a pixel.
\end{itemize}
Moreover, the shape of effective beams varies across the sky. Our aim is to estimate the effect of asymmetric effective beams and account for it in data analysis.

One of the immediate advantage of effective beams is that the point spread functions, the transpose of effective beams, become easily derivable {\em at every pixel}, which is otherwise non-trivial. For instance, in order to compute PSF at neighboring pixels using point source convolution methods, convolutions must be performed separately for each pixel to prevent overlap between PSFs. Limitation of the effective beams method is that the PSFs are centered at the pixel centers, though it may not be an issue in (pixelized) map based applications.

As mentioned before, Monte Carlo simulations are extremely important for studying systematic effects. Consider a sky map observed with certain detector response and scanning. To study the systematics in this map, one would simulate a number of model input maps and convolve them with the same detector response and scanning to estimate the average effect on these simulated maps. This can then be used to draw inferences on the map observed by the instrument. Since at each simulation scanning and beam are the same, huge amount of computation can be saved by pre-computing the effective beams, which include the effect of beam and scan once and for all for the whole set of simulations.

Once the effective beam is computed, it is rather straightforward and notably fast to make simulated ``observed'' maps. This is indeed the main goal of this paper - to show that the effective beam convolution is fast and accurate. In the end, we also provide some insights on the observed systematic effects directly by studying the properties of the effective beams.

The cost of effective beam computation scales as $\ntod \times \nbeam$, where $\ntod$ is the total number of samples, and $\nbeam$ is the number of pixels required to adequately describe the beam, and cost of convolution scales as $\npix \times \nbeam$, where $\npix$ is the total number of equal area pixels in the map. In general, $\nbeam = \npix$, as the far side lobe of the detector response can span the whole sky. In practice, however, the beam is negligible at most of the pixels; in fact, for CMB experiments of interest, non-negligible part of a main beam is very small and localized\footnote{Here we treat the far side lobes separately---they can be better modeled in the spherical harmonic domain at a much lower angular resolution~\citep{WandeltGorski-00}.}, leading to a big saving in computation and Input-Output (I/O) handling.

The finite size of the instrumental beam response causes exponential damping of angular power spectra at high multipoles, where noise becomes dominant. As a result, a CMB map observed with a certain Full Width at Half Maximum (FWHM), ${\rm FWHM}_{\rm beam}$, has usable information content typically up to a multipole of $\ell_{\rm max} \sim 4/{\rm FWHM}_{\rm beam}$. Also, a CMB map is pixelated in such a way that the number of pixels is minimized while retaining all non-negligible information. According to the HEALPix guidelines, this can be achieved by choosing $\npix \sim 200/{\rm FWHM}_{\rm beam}^2$. HEALPix divides the sky into $\npix = 12 \, \nside^2$ pixels, where $\nside$ is an even power of $2$ \citep{HEALPix}. Thus the above choice of pixelization translates to, $\nside \sim \ell_{\rm max} = 4/{\rm FWHM}_{\rm beam}$. As an example, a map observed with $\sim 30'$ beam is usually tessellated with $\npix = 3145728$ HEALPix pixels (i.e., $\nside = 512$). Then, if we assume that the beam is significant only up to an axial distance of $r$ times FWHM, the number of pixels required to describe  the beam would be
\begin{eqnarray}
&&\nbeam \ \approx \ \frac{\npix}{4\pi} \times {\rm area} \\
&&\qquad \sim \frac{200}{4\pi \, {\rm FWHM}_{\rm beam}^2} \pi \, (r \, {\rm FWHM}_{\rm beam})^2 \nonumber\\
&& \qquad  \sim \ 50 \, r^2 \, ,
\end{eqnarray}
which means that, if the beam is significant up to a {\em diameter} of four times FWHM, the number of pixels needed to describe it with adequate  accuracy is $\sim 200$. Note that, a symmetric Gaussian beam falls to $2^{-16} \approx 0.0015\%$ ($-48$dB) of its peak value at this diameter, and, in section~\ref{subsec:arealoss}, we explicitly show that the net power (effective area) outside this diameter is $\sim 0.001\%$ of the total beam power.

We compare the computational efficiency of pixel space convolution with the fast harmonic space convolution algorithm ``Total Convolution'' \citep{WandeltGorski-00, ChallinorEtAl-00}. Total Convolution uses a Fast Fourier Transform (FFT) on the sphere and convolution in the conjugate space to generate a ``Total Convolution Data Cube'' (TCDC)---sets of convolved sky maps for an equi-spaced grid of position and orientation coordinates of the beam. Total Convolution is theoretically exact at those grid points for band limited beam functions. Even though Total Convolution itself does not incorporate satellite scan strategy, the algorithm sets the limit on computation cost and accuracy that can be achieved by harmonic space simulation methods. Below we provide approximate theoretically estimated computation cost scaling of effective beam and total convolution; later we provide numerical comparison of accuracy. Note, however, that effective beam method also includes scan strategy (in the precomputation stage), to do the same in harmonic space one goes through the whole scan strategy and interpolates a TCDC, adding a computation cost which scales poorly due to machine I/O.

Computation cost of Total Convolution scales as $\sim \ell^3_{\rm max} \, m_{\rm max}$~\citep{WandeltGorski-00, ChallinorEtAl-00, conviqt}, where $\ell_{\rm max}$ is the highest multipole moment that contains non-negligible information about the anisotropies in the map and $m_{\rm max}$ is the highest azimuthal multipole needed to adequately describe beam asymmetry. In order to maintain minimum accuracy  in convolution, one needs to use $\ell_{\rm max} \sim \sqrt{\npix/3} \ ( = 2 \, \nside)$ and $m_{\rm max} \gtrsim 10$, which means that the computational cost would scale as $\gtrsim \npix^{3/2}/\sqrt{3}$. Since convolution in pixel space scales as $\sim \npix \, \nbeam$, pixel based convolution would be faster than total convolution as long as $\npix >  3 \times  \nbeam^2 \sim 10^5$, as $\nbeam$ was shown to be few hundred for a reasonable beam cutoff radius.

As effective beam computation and convolution are the focus of the paper, the rest of the paper provides details of computation and accuracy analysis considering the main beam only. In addition, though our method and implementation can handle any arbitrary beam shape, in order to reliably compare our results with existing Planck simulations we consider only elliptic Gaussian beam for numerical results.

\section {Mathematics for Effective Beams}
\label{sec:math}


Computation of effective beams involves large number of mathematical operations. Hence, it is important to formulate it in a computationally minimalist manner. In this section we list the formulae we use in our code. Before we proceed further we define some quantities which will be used throughout the paper:

\begin{itemize}

\item  Effective beam for pointing pixel $i$ and beam pixel $j$: $\effBeam_{ij} \equiv \bm{\effBeam}$.

\item Pointing matrix for pixel $p$ and sample (or time) $t$:
\begin{equation}
\mathbf{A} \ \equiv \ A_{tp} \ = \
\left\{\begin{array}{cl}
1 \, , \ & \ \textrm{if sample lies in pixel}\\
0 \, , \ & \ \textrm{otherwise}
\end{array}\right.  \nonumber
\end{equation}

\item Measured signal or Time Ordered Data (TOD) at sample $t$: $d_t \equiv \mathbf{d}$.

\item Pointing direction at a sample $t$: $\urad(t) \equiv [\theta(t), \phi(t)]$.

\item Polarization angle, the angle between the polarization axis of the detector and the local meridian: $\psi_t \equiv \psi(t)$.\\
(for unpolarized detectors, polarization axis is the $x$-axis of the fiducial detector frame)

\item The set of three pointing angles that define the instantaneous geometry of each detector $\upoint_t \equiv [\theta(t),\phi(t), \psi(t)]$

\item Angle between polarization axis and the $x$-axis of the fiducial detector frame: $\psi_{\rm pol}$.

\item Angle between the major axis of an elliptical beam (or any preferred axis relative to which the beam profile has been defined) and the $x$-axis of the fiducial detector frame: $\psi_{\rm ell}$.

\item Detector intrinsic beam function at a direction $\urad_b \equiv [\theta_b, \phi_b]$ with the pointing angles $\upoint$: $b(\urad_b, \upoint)$.

\item Direction vector corresponding to pixel number $p$: $\ncap_p$.

\end{itemize}

Note that the polarization angle is being used to describe the orientation of the detector without any loss of generality. Since the polarization axis of the detector is fixed to the fiducial detector frame, any other orientation angle in this frame, e.g., the orientation angle of the major axis of an elliptic beam, $\psi_{\rm ell}$, can be derived by adjusting the appropriate offset angles. For an unpolarized detector, the angle between the polarization axis and the fiducial detector frame is taken as zero.

The detector intrinsic beam function is generally given in a 2-D coordinate system on the beam plane, with the origin at the beam center and the $x$-axis conveniently chosen---e.g., for an elliptic Gaussian beam the major axis is chosen as the $x$-axis. Here we consider the beam to be defined on the tangent plane, in general the beam could as well be specified in the ``curved'' polar coordinates at the North pole, then we could take the $\phi = 0$ direction as the $x$-axis.

\subsection{Temperature-Only Effective Beam}
\label{subsec:TonlyEB}

By definition, effective beams are the objects whose convolution with the true CMB sky should produce the observed sky map. So we write the expression for the observed CMB map (without noise) in a convolution form and then define effective beams. 

Each observed sample is a convolution of beam and the CMB temperature anisotropy field $T(\urad)$,
\begin{eqnarray}
d_t &=& \int \d \Omega_{\urad_b} \, b(\urad_b,\upoint_t) \, T(\urad_b) \\
&\approx& \ \DelOm \sum_j b(\urad_j,\upoint_t) \, T_j \, . \label{eq:obsTTOD}
\end{eqnarray}
where  $\mathbf{T} \equiv T_p$ is the temperature map in pixel basis.
Typical mapmakers however ignore the effect of extended beam, as it is computationally prohibitive, and  it ``bins'' (averages) the observed data at each pixel (in the absence of correlated noise).  So the binned map is made assuming an infinitely narrow (Dirac $\delta$-function) beam, in the discrete pixel space, which is equivalent of setting the beam response to $1$ in the pointing pixel and $0$ outside (that is, a Kronecker $\delta$). Then the observed binned map $ \widetilde{\mathbf{T}} \equiv \widetilde{T}_p$ can be written as
\begin{equation}
\widetilde{T}_i \ :=\ \frac{\sum_{t} A_{ti} \, d_t}{\sum_{t} A_{ti}} \, .
\end{equation}
Here the numerator adds all the detected signal if the pointing direction falls in pixel $i$, and the denominator counts the number of observations in that pixel. However, since the beam is extended, the observed map is connected to the true underlying anisotropy map through the following relation [see Eq.~(\ref{eq:obsTTOD})]:
\begin{eqnarray}
\widetilde{T}_i &:=& \frac{\sum_{t} A_{ti} \, d_t}{\sum_{t} A_{ti}} = \frac{\sum_{t} A_{ti} \, \DelOm \sum_j b(\urad_j,\upoint_t) \, T_j}{\sum_{t} A_{it}} \nonumber\\
\label{eq:binnedT}\\
&=& \DelOm \sum_j \left[ \frac{\sum_{t} A_{ti} \, b(\urad_j,\upoint_t)}{\sum_{t} A_{ti}} \right] \, T_j \, .
\end{eqnarray}
Then, in order to satisfy the definition of effective beam,
\begin{equation}
 \widetilde{\mathbf{T}} \ =: \ \DelOm \, \bm{\effBeam} \cdot \mathbf{T},
\end{equation}
one must choose
\begin{equation}
\effBeam_{ij} \ = \ \frac{\sum_t A_{ti} \, b(\urad_j, \upoint_t)}{\sum_t A_{ti}}.
\label{eq:EBT}
\end{equation}
This is indeed the crude definition of effective beam stated in section~\ref{sec:EB}, which directly applies to the temperature-only case. In fact, the reason for arriving at the expression above in a formal way is to aid our understanding for the case of polarized detectors, where the definition is not straightforward.

Eq.~(\ref{eq:EBT}) is essentially what we compute for temperature only beams, the detailed method of how it is done in an efficient way is discussed in this section and section~\ref{sec:Algo}. 

It is clear from Eq.~(\ref{eq:EBT}) that for each sample $t$, the beam function $b$ must be evaluated at all the beam pixels $j$ by rotating the detector intrinsic beam function. We do this by transforming the pixel center coordinates to the coordinates in which the detector intrinsic beam function is specified. Note that these operations are the {\em most expensive mathematical operations} in the effective beam computation, as they are used about trillions times ($\ntod \times \nbeam$), so we take good care in writing them in efficient forms.

\begin{itemize}

\item First, we obtain the coordinates of the point in the (locally flat) orthogonal coordinates defined by the polar and azimuthal unit vectors ($\utheta,\uphi$) with the beam center ($\urad$) as the origin.

The radial, polar, and azimuthal unit vector in spherical polar coordinates are respectively given by:
\begin{eqnarray}
\urad &=& \sin\theta \cos\phi \, \icap \, + \,  \sin\theta \sin\phi \, \jcap \, + \, \cos\theta \kcap \nonumber\\
\utheta &=& \cos\theta \cos\phi \, \icap \, + \,  \cos\theta \sin\phi \, \jcap \, - \, \sin\theta \kcap \nonumber\\
\uphi &=& -\sin\phi \icap \, + \, \cos\phi \jcap \nonumber\\
\end{eqnarray}

Then, with respect to the beam centre, the coordinates ($\Delta x, \Delta y$) of a pixel at $\urad_b$ can be written as:
\begin{eqnarray}
\Delta x &:=& (\urad_b - \urad) \cdot \utheta\\
&=& \  \sin\theta_b \, \cos\theta \, \cos(\phi_b-\phi) \nonumber\\
&& \qquad - \  \sin\theta \, \cos\theta_b \label{eq:accDelx}\\
 \nonumber\\
\Delta y &:=&  (\urad_b - \urad) \cdot \uphi \\
&=& \sin\theta_b \, \sin(\phi_b - \phi) \, . \label{eq:accDely}
\end{eqnarray}

We also provide a less accurate but faster (by $\sim 10\%$ of the {\em whole} computation) set of formulae to get these coordinates. We have tested that these formulae provide enough accuracy for practical purposes.
\begin{eqnarray}
\Delta x &=& - \theta \ + \nonumber\\
&&\quad \left\{ \begin{array}{l}
\theta_b \, \cos(\phi_b - \phi) \, ,\\
\quad {\rm if} \ \theta \le \pi/2 \\
\pi - (\pi-\theta_b) \, \cos(\phi_b - \phi) \, ,\\
\quad {\rm otherwise}
\end{array} \right. \nonumber\\
\\
\Delta y &=& \ \sin\theta \ \times \nonumber\\
&&\quad \left\{\begin{array}{l}
\sin(\phi_b - \phi) \, ,\\
\quad {\rm if} \ |\phi_b - \phi| > 1^\circ \  {\rm and}\\
\qquad |\Delta x| < {\rm FWHM}_{\rm beam}\\
\phi_b - \phi \, ,\\
\quad {\rm otherwise} \, .
\end{array}\right. \nonumber\\
\end{eqnarray}

\item We then rotate the coordinates by the angle $\psi_b = \psi(t) - \psi_{\rm pol} + \psi_{\rm ell}$ to transform to the beam coordinate system:
\begin{eqnarray}
&&\left(\begin{array}{c}
\Delta x_{\rm beam}\\
\Delta y_{\rm beam}
\end{array}\right)
\ = \nonumber\\
&&\qquad \left[\begin{array}{rr}
\cos\psi_b & \sin\psi_b\\
-\sin\psi_b & \cos\psi_b
\end{array}\right]
\left(\begin{array}{c}
\Delta x\\
\Delta y
\end{array}\right). \nonumber\\
\end{eqnarray}

\item The beam function is a known input in the new $\Delta x_{\rm beam}, \Delta y_{\rm beam}$ coordinates.  For example, if the intrinsic beam is elliptical Gaussian with major and minor axes $\sigma_a, \sigma_b$ respectively, the value of the rotated beam function at $\urad_b$ will be
$$\frac{1}{2 \pi \sigma_a \sigma_b} \exp\left[-\frac{\Delta x_{\rm beam}^2}{2 \, \sigma_a^2} - \frac{\Delta y_{\rm beam}^2}{2 \, \sigma_b^2}\right].$$

\end{itemize}

In this paper all the numerical exercises have been done assuming elliptical Gaussian beam shape with parameters read from a database of Planck focal plane parameters and the accurate formulae for beam coordinates, Eq.~(\ref{eq:accDelx}) and Eq.~(\ref{eq:accDely}), have been used.

\subsection{Polarized Effective Beams}

To define effective beams for polarized detectors we again formally express the output of a map-maker in the noiseless case. This expression is then generalized to the extended beam case and the polarized effective beam is defined. In case the beam is significantly extended, we also need to consider the coordinate changes within the beam (as the local meridians---the polarization reference axes---at different pixels may become significantly different). We supply the correction formulae in the end of this subsection, though for Planck main beams we achieved adequate accuracy without using this correction. 

\subsubsection{Polarized binned map}

The binning process is independent of the beam size. However, if the beam has infinite resolution ($\delta$-function) the binned map should be identical to the true map in the noiseless case. So we derive the formula for making binned polarized maps starting with such an idealized case, with the view that the procedure will be valid for any beam.

A detector in a CMB experiment records only one number for each sample, a linear combination of three Stokes parameters $I(\urad), Q(\urad), U(\urad)$. For the idealized detector, one can write
\begin{eqnarray}
d_t &=& w_{1}(\psi_t) \, I(\urad(t)) \ + \\
&& \qquad  w_{2}(\psi_t) \, Q(\urad(t)) \ + \ w_{3}(\psi_t) \, U(\urad(t)) \, , \nonumber
\end{eqnarray}
where $\mathbf{w}_t \equiv [w_{1}(\psi_t), w_{2}(\psi_t), w_{3}(\psi_t)]$ is the weight vector at sample $t$.
The above equation can be written in a discrete form using the pointing matrix $A_{tp}$ as
\begin{equation}
d_t = \sum_{p} A_{tp} \mathbf{w}^T_t  \left(\begin{array}{c}I\\Q\\U\end{array}\right)_{p} \, .
\label{eq:polTOD}
\end{equation}
For the polarization sensitive instruments used in the Planck satellite, it is a good approximation to write the weight factors as,
\begin{equation}
\mathbf{w}_t  := \left( \begin{array}{c}
1 \\ \gamma \, \cos(2 \psi_t) \\ \gamma \, \sin(2 \psi_t)
\end{array}\right),
\end{equation}
where $\gamma := (1-\epsilon)/(1+\epsilon)$ is the polarization efficiency, $\epsilon$ being the {\em cross-polar leakage}.

Since the detector provides one number for each sample, but one needs to estimate three numbers ($I,Q,U$) for each pixel, there must be at least three observations per pixel and the set of linear equations must be solved. Ignoring the issues of degeneracy (not enough independent detector configurations), below we describe how one can solve the equations to make a binned map.

The goal of making a binned map is that, if the beam was indeed vary narrow and there was no noise, one should get the true sky map as the result. We repeat the procedure presented in \citet{JonesEtAl-06}. Multiply Eq.~(\ref{eq:polTOD}) through by $\mathbf w_t$ and sum over the samples in a pixel (enumerated by the pointing matrix $A$), to obtain
\begin{equation}
  \sum_t A_{tp} \mathbf{w}_t d_t
= \sum_t  \sum_{p'} A_{tp} A_{tp'}  \mathbf{w}_t \mathbf{w}^T_t  \left(\begin{array}{c}I\\Q\\U\end{array}\right)_{p'}. 
\end{equation}
Because the pointing matrix has only one unit entry per sample, the sum over $p'$ collapses to:
\begin{equation}
  \sum_t A_{tp} \mathbf{w}_t  d_t = \mathbf{H}_p  \left(\begin{array}{c}I\\Q\\U\end{array}\right)_{p} \, .
\end{equation}
On the left hand side we have binned weighted TOD and on the right we have the true sky times the $3 \times 3$ polarized ``number of observation matrix'' at pixel $p$ 
\begin{equation}
  \mathbf{H}_p \ := \ \sum_t A_{tp} \mathbf{w}_t \mathbf{w}^T_t \, .
\end{equation}
To recover the sky,  multiply the binned weighted TOD by the inverse of number of observation matrix. So the polarized binned map is given by
\begin{eqnarray}
&&\left(\ \widetilde{I} \quad \widetilde{Q} \quad \widetilde{U} \ \right)_p^T \ = \ \mathbf{H}_p^{-1}  \,  \sum_t A_{tp} \mathbf{w}_t  d_t \nonumber\\
&& \qquad = \ \left[ \sum_t A_{tp} \mathbf{w}_t \mathbf{w}^T_t \right]^{-1} \sum_t A_{tp} \mathbf{w}_t  d_t \, . \nonumber\\
\label{eq:binMap}
\end{eqnarray}
As mentioned above, this is the general binning formula for any beam size. Also note that, not surprisingly, it is similar to the formula for the temperature only case [Eq.~(\ref{eq:binnedT})].

\subsubsection{Polarized effective beams}

We can now calculate the effect of binning on a data set observed with a realistic beam of finite size and thereby arrive at the formula for polarized effective beams. We follow the same path as we did for temperature only beams.

Following \citet{JonesEtAl-06}, the observed signal at sample $t$ is simply the integral of the contraction:
\begin{eqnarray}
d_t &=& \int \d\Omega_{\urad_b} \, b(\urad_b,\upoint_t) \, \mathbf{W}^T(\urad_b,\upoint_t) \cdot (I \quad Q \quad U)^T_p \nonumber\\\\
&\approx& \DelOm \sum_{p} b(\ncap_p,\upoint_t) \, \mathbf{W}^T(\ncap_p, \upoint_t) \cdot (I \quad Q \quad U)_{p}^T \, , \nonumber\\
\end{eqnarray}
where $\mathbf{W}(\urad_b,\upoint_t)$ is the polarization weight vector for general extended beam,
\begin{equation}
\mathbf{W}(\urad_b, \upoint_t)  := \left( \begin{array}{c}
1 \\ \gamma \, \mathcal{P}(\urad_b,\upoint_t) \cos(2 \psi_t) \\ \gamma \, \mathcal{P}(\urad_b,\upoint_t) \sin(2 \psi_t)
\end{array}\right),
\end{equation}
where $\mathcal{P}(\urad_b,\upoint)$ is the normalized beam response function, defined as
\begin{equation}
\mathcal{P}(\urad_b,\mathbf{0}) \ := \ \frac{P_\parallel(\urad_b) - P_\perp(\urad_b)}{P_\parallel(\urad_b) + P_\perp(\urad_b)} \, ,
\end{equation}
expressed  in terms of the co- and cross- polar beam response functions $P_\parallel(\urad_b)$ and $P_\perp(\urad_b)$ respectively. $\mathcal{P}(\urad_b,\upoint)$ can be obtained for any given pointing $\upoint$ following the same procedure as that for the beam profile $b(\urad_b,\upoint)$ described explicitly for temperature only beams.

On substitution into the mapmaking equation [Eq.~(\ref{eq:binMap})], the observed (convolved) sky map becomes:
\begin{eqnarray}
&&\left(\begin{array}{c} \widetilde{I}\\ \widetilde{Q}\\ \widetilde{U}\end{array}\right)_i \ = \  \mathbf{H}_i^{-1}  \,  \sum_t A_{ti} \, \mathbf{w}_t  \times\\
&& \qquad \DelOm \sum_j b(\urad_j,\upoint_t) \, \mathbf{W}^T(\ncap_j,\upoint_t)  \cdot \left(\begin{array}{c} I\\ Q\\ U\end{array}\right)_j \, . \nonumber
\end{eqnarray}
Thus, in order to satisfy the definition of polarized effective beam,
\begin{equation}
( \widetilde{I} \quad \widetilde{Q} \quad \widetilde{U})_i^T \ =: \ \DelOm \sum_j \effBeamPol_{ij} \cdot (I \quad Q \quad U)_j^T,
\end{equation}
one must choose,
\begin{eqnarray}
\effBeamPol_{ij} &=&  \left[ \sum_t A_{tp} \mathbf{w}_t \mathbf{w}^T_t \right]^{-1} \\
&& \qquad \sum_t A_{ti} \, b(\urad_j,\upoint_t) \, \mathbf{w}_t  \mathbf{W}^T(\ncap_j,\upoint_t) \, . \nonumber
\end{eqnarray}
The effective beam for each pointing pixel $i$ now has two parts. There is a $3 \times 3$ symmetric matrix $\mathbf{H}_i$, which is a generalized version of number of observation at that pixel. Then, for each neighboring pixel $j$ there is a $3 \times 3$ matrix $\sum_t A_{ti} \, b(\urad_j,\upoint_t) \, \mathbf{w}_t  \mathbf{W}^T(\ncap_j,\upoint_t)$, which encapsulates the beam profile and need not be a symmetric matrix in general. Computation of this quantity  can be done in a similar way as in the temperature only case, first computing the coordinates of neighboring pixels with respect to the beam center and then evaluating the polarized beam profiles $b(\urad_j,\upoint_t) \mathbf{W}^T(\ncap_j,\upoint_t)$ using the measured/modeled intrinsic detector response functions.

The cross polar beam is normally much smaller than the co-polar beam and, in practice, the co-polar beam might be the only response which could be accurately measured. These facts lead to a reasonable assumption that $\mathcal{P}(\urad_b) \approx 1$, so that, $\mathbf{W}(\urad_b,\upoint_t) = \mathbf{w}_t$, which, in turn, simplifies the computation of effective beams,
\begin{equation}
\effBeamPol_{ij} =  \left[ \sum_t A_{tp} \mathbf{w}_t \mathbf{w}^T_t \right]^{-1} \sum_t A_{ti} \, b(\urad_j,\upoint_t) \, \mathbf{w}_t  \mathbf{w}^T_t ,
\end{equation}
as the beam profile for all the polarization becomes the same, except for weight factors (not dependent on $\urad_j$), and makes all the component matrices symmetric.  Note that, if the effective beam matrices are not symmetric\footnote{``Polarized effective beam matrices are symmetric'' signifies that the $3 \times 3$ matrix for representing polarized effective beams is a symmetric matrix, it should not be confused with an effective beam constructed from symmetric detector beams.}, computational storage requirement increases by $50\%$ ($9$ instead of $6$ components for each matrix). Computation time is not much affected by this modification.

{\tt FEBeCoP} does include the ability to compute effective beam matrices of both kinds, symmetric or otherwise. However, the numerical results and comparisons presented in this paper are for the possibly more practically achievable scenario of $\mathcal{P}(\urad_b) = 1$. So, unless otherwise specified, the matrices representing polarized effective beams computed for accuracy and performance analysis of this method would be symmetric.

\subsubsection{Transformation from beam frame to sky frame}

If the beam is significantly extended, the pixels which are   far (that is on very different longitudes) would suffer from an extra polarization rotation effect, this is because the far away longitudes are no longer close to parallel (except when they are very close to the equator), so the frame of polarization also rotates. The mathematics to account for that extra rotation is presented below.

\begin{figure}
\centering
\includegraphics[width=0.45\textwidth]{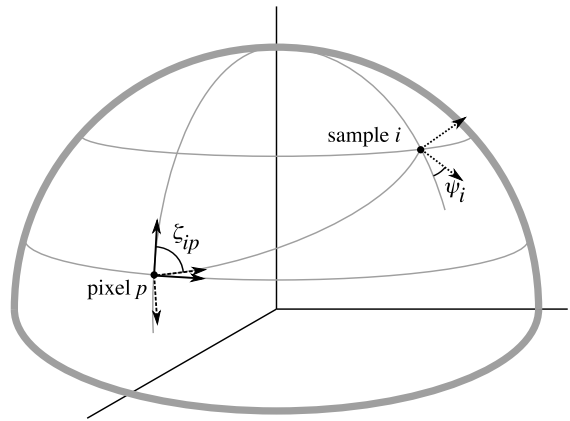}
\caption{Orientation of the detector for sample $i$ after rotation by Euler angles $\{\phi_i,\theta_i,\psi_i\}$, and the orthonormal basis $(\hat \phi, \hat \theta)$ at pixel $p$ in both the sky frame (where $\hat \theta$ points along the great circle to the pole ) and the beam frame (where $\hat \theta$ points to the detector).  The angle between the polarization bases is labeled $\zeta_{ip}$.} \label{fig:geom}
\end{figure}

If the detector orientation at sample $t$ is represented by Euler rotation matrix $\mathbf{D}(\phi_t,\theta_t,\psi_t)$, then the polarization field for the beam in pixel $j$ is evaluated at a direction $\mathbf{D}^{-1}(\phi_t,\theta_t,\psi_t) \cdot \ncap_j$ in the beam frame. The orthonormal polarization bases differ between frames by a rotation.  Stoke $I$ is a scalar, unchanged by the rotation.  To transform $(Q,U)$, we need to know the angle between the polarization bases in the sky and beam frame (Figure~\ref{fig:geom}). 
Imagining the spherical triangle between the North pole, the position of sample $i$, and pixel $p$, we may deduce the angle between the orthonormal bases at the pixel location, denoted $\zeta_{ip}$, with two applications of the law of haversines,
\begin{eqnarray}
&& {\rm haversin} (\Delta \Phi_{ip}) \ = \ {\rm haversin} (\theta_i -\theta_p) \ + \nonumber \\
&& \qquad \sin(\theta_i) \sin(\theta_p) {\rm haversin} (\phi_i -\phi_p) \\
&& {\rm haversin} (\theta_i) \ =\ \quad {\rm haversin} (\theta_p - \Delta \Phi_{ip}) \ + \nonumber\\
&&  \qquad \sin(\theta_p) \sin(\Delta \Phi_{ip}) {\rm haversin} (\zeta_{ip}),
\end{eqnarray}
where ${\rm haversin}(\theta) = \sin^2(\theta/2)$.  The first equation solves for $\Delta \Phi_{ip}$, representing the great circle distance from the pixel to the sample position, which is required in the second equations to solve for $\zeta_{ip}$.

This extra rotation would lead to a transformation of the observed sky in the following way,
\begin{equation}
\left(\begin{array}{c}Q' \\ U'\end{array}\right) \ = \
\left[\begin{array}{rr}
\cos 2\zeta_{tp} & \sin 2\zeta_{tp} \\
-\sin 2\zeta_{tp} & \cos 2\zeta_{tp}
\end{array}\right]
\left(\begin{array}{c}Q \\ U\end{array}\right).
\end{equation}
This would change the beam weight factor in the expression for effective beams to $\mathbf{W}(\ncap_p,\upoint_t) \rightarrow  \mathbf{W}'(\ncap_p, \upoint_t) = \mathbf{W}(\psi_t + \zeta_{tp},\upoint_t)$ and the expression would become
\begin{eqnarray}
\effBeamPol_{ij} &=& \left[ \sum_t A_{tp} \mathbf{w}_t \mathbf{w}^T_t \right]^{-1}\\
&&\qquad \sum_t A_{ti} \, b(\urad_j,\upoint_t) \,\mathbf{w}_t   \mathbf{W}'(\ncap_j, \upoint_t) \, . \nonumber
\end{eqnarray}
It is easy to see that the effective beam matrix at each pixel is no longer symmetric in this case, even if the cross-polar beam is ignored.

\section{Theoretical Accuracy}
\label{sec:TC}

Effective beams defined over a grid could have quadrature accuracy issues if the pixelization was not fine enough. Also, the radial cut off we impose on the beam leads to some information loss. In this section we study the truncation error in integrating a 2-D Gaussian over a HEALPix grid to determine the minimum value of the cut-off radius and also show that quadrature error is sub-dominant here. Then, by comparing effective beam convolved CMB maps using Total Convolution for a toy scan model, we show that the default map pixel resolution (discussed in section~\ref{sec:EB}) provides sufficient accuracy.

\subsection{Effective Area Loss of a Truncated Gaussian}
\label{subsec:arealoss}

By definition, the surface integral of a 2-D Gaussian over an infinite plane is unity. However, if we cut off a zero mean Gaussian with variance $\sigma^2$ at $r_{\rm max} = n_{\rm FWHM} \, \times \, {\rm FWHM} \ = \ 2 \, n_{\rm FWHM} \,\sqrt{2\ln{2}} \, \sigma$, the ``area loss'' turns out to be
\begin{eqnarray}
\epsilon  &:=& 1 - \frac{1}{\sigma^2} \int_0^{r_{\rm max}} \d r \, r \, e^{-\frac{r^2}{2\sigma^2}} \ = \ e^{-\frac{r_{\rm max}^2}{2\sigma^2}} \nonumber\\
&=& 2^{- 4 \, n_{\rm FWHM}^2} \, .
\end{eqnarray}
The order of magnitude error is thus
\begin{equation}
\log_{10}{\epsilon} \ = \ 4 \, n_{\rm FWHM}^2 \, \log_{10}{2} \ \approx \ 1.2 \, n_{\rm FWHM}^2 \, .
\end{equation}
That is, if we choose the cutoff at $n_{\rm FWHM} = 2$, the area loss will be $\sim 10^{-5}$, which should provide enough accuracy for convolution.
It is worth noting that in the context of CMB maps the error is actually the difference between convolved values with and without cutoff, which is dependent on the underlying CMB map. Since the maps have zero expectation, the expected error would also be zero, but the fractional RMS error would be the same as the effective area loss.

In practice convolution is done in discrete space. To confirm that the above analytical estimate still holds and the usual choice of pixel resolution does not add any significant quadrature error, we do the exercise numerically. We compute the area loss for a Gaussian with FWHM $\sim 32'$ (Planck LFI 30GHz beam size) using two sets of beam cut-off radius, $n_{\rm FWHM} = 2$ and $n_{\rm FWHM} = 4$, and pixelization, $\npix = 12 \times 512^2$ and $\npix = 12 \times 2048^2$.

\begin{figure}
\centering
\includegraphics[width=0.45\textwidth]{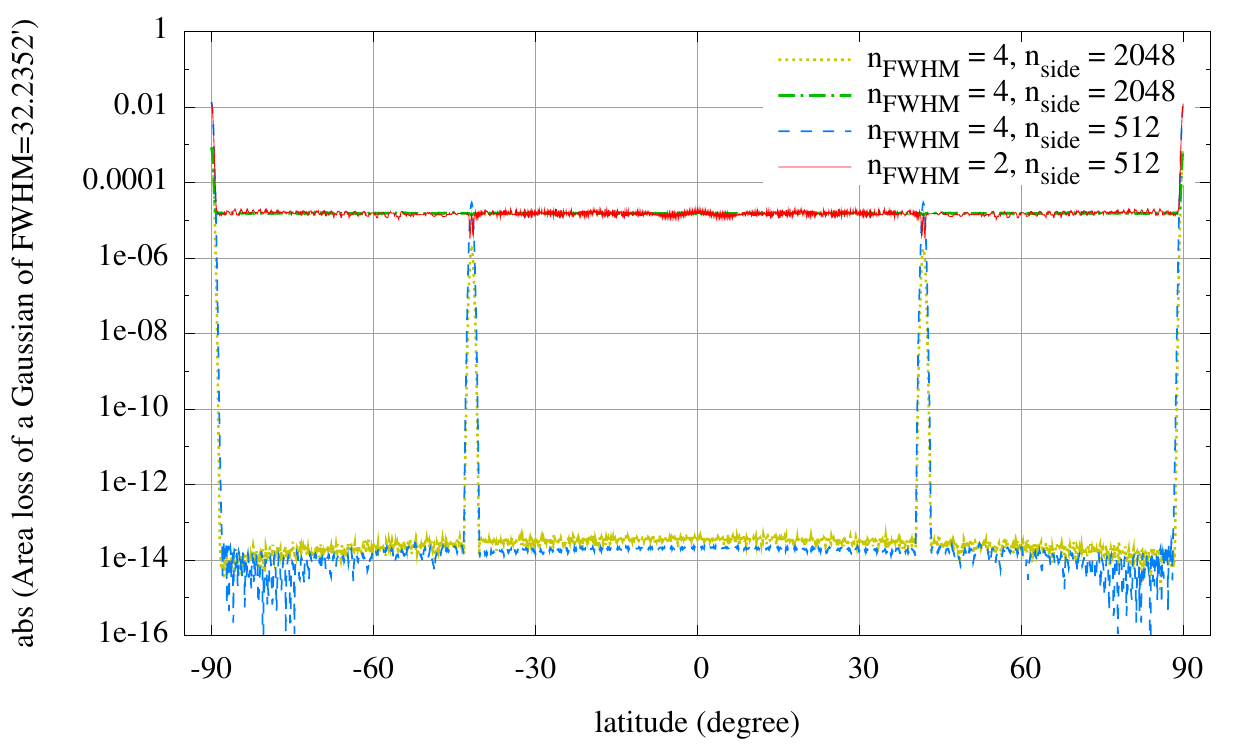}
\caption{A symmetric Gaussian beam of FWHM=32.2352'  was integrated in the pixel space for two different HEALPix pixel resolution as well as two different radial cut-offs (expressed in the units of FWHM of the Gaussian). In the ideal case the result of the integration---the ``area'' of the Gaussian---should be unity. The deviation of the numerical results from unity, the ``area loss'', is shown in the plot. As expected, the area loss is mainly dependent on the cut-off radius and very weakly dependent on the quadrature error due to pixelization. The sharp features in the plots are related to the pixelization scheme (HEALPix). The point to note here is that, for a cut-off radius of $n_{\rm FWHM} = 2$ and $\nside = 512$ the errors are at the level of single precision floating point accuracy, which we consider sufficient for the target accuracy of map simulation.}
\label{fig:areaLoss}
\end{figure}
Figure~\ref{fig:areaLoss} shows plots of area loss for the four cases as functions of latitude (the evaluation was done along the meridian without any loss of generality). Clearly, area loss depends on the cut-off radius and can not be improved by increasing $\npix$ to higher than the usual value for the given beam size. There is some quadrature error near the poles, but we can ignore that as the solid angle which is affected by this is very small (only few pixels). There is a sharp features near each HEALPix polar circle for higher cutoff radius, only if the target accuracy is better than $10^{-5}$. Our default choice would be $n_{\rm FWHM} = 2$, which provides adequate accuracy without bringing in any significant quadrature issues.

\subsection{Comparison with Total Convolution}

Next we compare effective beam convolved maps for toy models of beam orientations with Total Convolution Data Cubes for different choices of pixel resolution. TCDC are theoretically exact for band limited beam functions on an equi-spaced grid of beam location and orientation. We used those values of orientation in the effective beam computation and convolved with the same map (in pixel basis) from which the TCDC were derived. TCDC are computed at equi-spaced intervals of RA and dec, so the beam centers do not coincide with the HEALPix pixel centers. This is similar to the real scans, which need not pass right through the pixel centers. One small approximation is unavoidable here, a Gaussian beam is not purely band limited, specially for asymmetric beam there is a cut-off on the azimuthal index `$m$'. This error is not severe unless the accuracy requirement is very stringent. In fact, this error adds to true differences making the test even more strict for the pixel space convolution.

\begin{figure}
\centering
\includegraphics[width=0.475\textwidth]{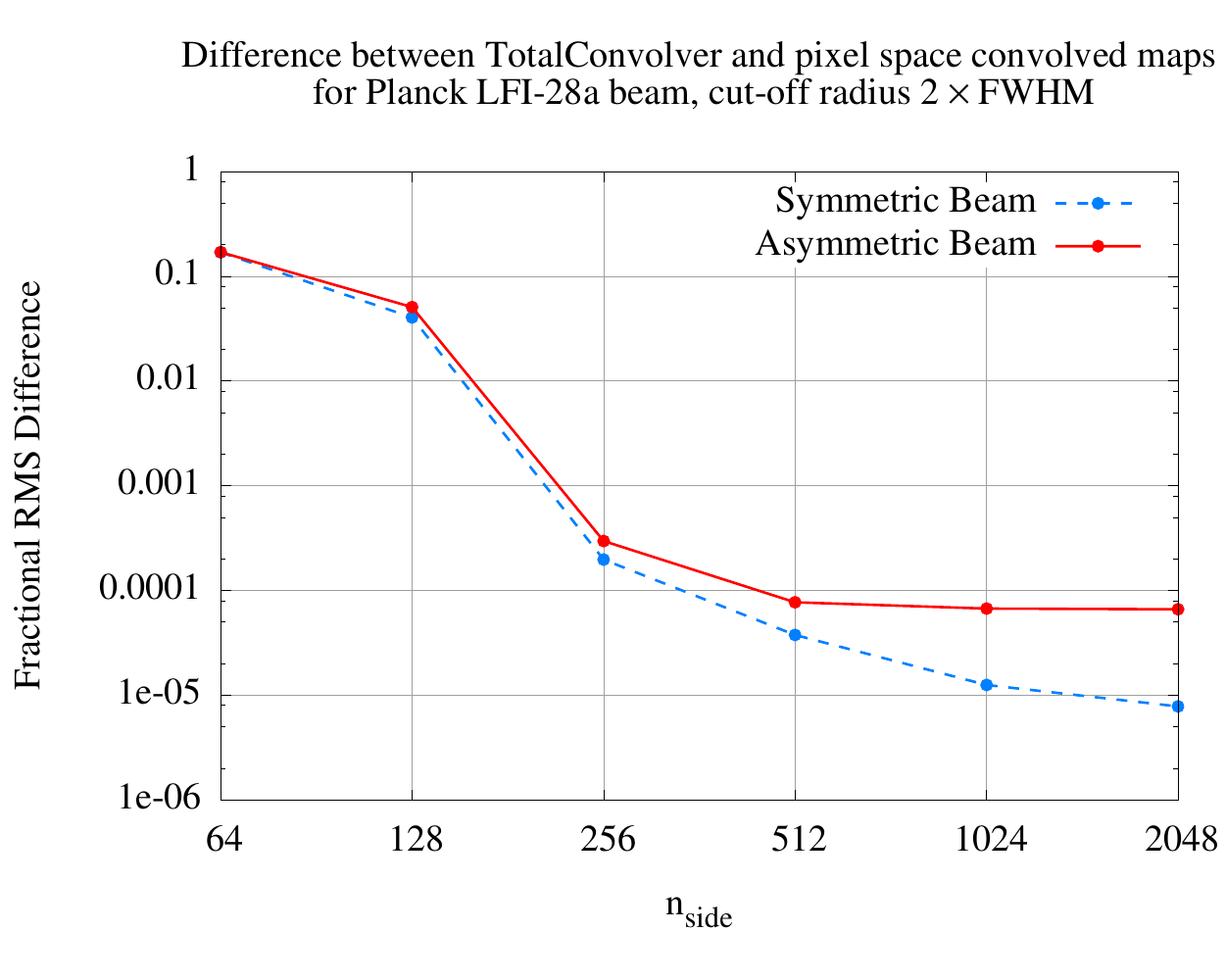}
\caption{Difference between Total Convolution maps (data cube slices) and pixel space convolved maps for different HEALPix $\nside$ is shown in this figure for symmetric (dashed) and asymmetric (solid) beams of FWHM $\sim 32'$ and ellipticity $\sim 1.4$. Total Convolution is exact on an equi-spaced ($\theta,\phi$) grid for band limited functions. The comparisons were performed on this grid to access how far pixel space convolution is from the exact theoretical convolution and to determine the optimal $\nside$ for effective beam computation. For each HEALPix pixel, the TCDC value at the closest grid point was used for comparison, no interpolation was performed.}
\label{fig:compEBTC}
\end{figure}
Figure~\ref{fig:compEBTC} shows that for a $\sim 32'$ beam, the natural choice of pixel resolution $\nside = 512$ provides very good accuracy and further increase in resolution does not improve it by any significant amount. At low $\nside$ the beam is very poorly sampled, which leads to orders of magnitude better accuracy for each step of $\nside$ improvement. However, beyond $\nside=512$ the amount of extra information added is low, so the accuracy improvement is marginal and is also limited by single precision floating point accuracy. So, we may conclude that {\em the usual choice of pixelization is the optimum choice}.

\section{Computation Algorithm and Implementation}
\label{sec:Algo}

 In the previous sections we have presented the theoretical statement of the problem and error analysis. The practical implementation of the method is a bigger challenge due to the organizational complexities of efficient distributed computation. This section outlines the possible routes of computation, their relative merits and demerits and some final outcomes.

\subsection{Quantities to Compute}

The definitions provided in section~\ref{sec:EB} suggest that in order to evaluate and store the effective beams, the following quantities have to be computed for each pixel:
\begin{itemize}

\item The list of pixels $\mathbf{p}$ that are within the beam cut-off radius from the pixel center. Ideally one should find this list with respect to the sample beam centers within the pixel, but the difference would be negligibly small and cost large amount of computation, as it will have to be done at each sample. Instead, the list is computed in the very beginning with respect to the pixel centers.
The number of element in the list, $\nlist$, is slightly different for each pointing pixel.

\item The number of observation $\nhits$ at a given pixel or, for polarized beams, the weight matrix $\mathbf{H}$ (whose first component is $\nhits$). Note that effective beams are computed in a way that they are always normalized (divided by $\nhits$) as they are developed.  That is, only the fractional changes from the existing beams are added.  This prevents large number arithmetic (leading to numerical errors) near the ecliptic poles where, for Planck, $\nhits$ is higher by $3$-$4$ orders of magnitude.

\item The effective beam $\effBeam_{ij}$ at the neighboring pixels for temperature only beam and the effective beam matrix $\effBeamPol_{ij}$ for each neighboring pixel for polarized beams. 

\end{itemize}

In addition, there can be intermediate quantities that need to be pre-computed for convenient execution of the main steps, this intermediate quantities can also consume significant amount of computation. Details of these implementation dependent steps are provided in the following sections.

\subsection {Algorithm}

From the description provided in section~\ref{sec:EB} it follows that there are two obvious ways to compute the effective beams, using a time-domain or pixel-domain parallelization scheme.  Their difference is purely computational, the challenge being to manage the large effective beam data object and divide computation efficiently.  We explored both approaches while developing our effective beam pipeline, settling on the pixel domain as our preferred method of parallelization.

\subsubsection{Time domain parallelization}

Time domain parallelization means different portions of the scan are assigned to different processes. As the scan progresses, more and more pixels are assigned to each process and memory is allocated progressively.  To avoid memory overflow, after certain predetermined interval, beams are written to disk---if for some pixels beams already existed on disk, they are read, new beams are added and rewritten to disk.

\begin{itemize}

\item {\bf Advantages}: Simplistic implementation (though the memory management process is complicated)

\item {\bf Disadvantages}: I/O intensive in different ways. Disk I/O is many times the total size of beams, as partial beams are read and written many times. Moreover, since one pixel can appear in different processes, the total beam (data) size is about three times bigger than the true size for one detector, leading to a massive loss of memory, disk space and overall I/O overhead; handling multiple detectors at a time is practically impossible, as rings would become ``thicker'' due to the angular distance between detectors, requiring much more memory for each ring; it is not easy to locate a pixel in the beam files which makes it very difficult to load beams for a certain set of pixels in memory - hence the beam files have to be read from disk for every convolution; there is no easy way to predetermine memory usage by a certain process, so memory efficient parallelization is practically impossible. Finally, each process requires the satellite pointing information simultaneously, which in turn requires reading the same satellite pointing file---this is a big bottleneck and increasing the number of processes does not necessarily speed up computation.

\end{itemize}

Due to the long list of disadvantages above, this method was not very useful for efficient computation of effective beams, especially for polarization. Though the simplicity of the implementation helped us making initial assessment of accuracy, computational requirement and feasibility of the method.

\subsubsection{Pixel domain parallelization}

Pixel domain parallelization means different pixels are assigned to different processes, though the division of pixels need not be equal as we explain below.

Since the satellite scan path is efficiently obtained in the time domain with contiguous stretch of samples, we need to go through an intermediate step of  binning the detector angles in pixels---for each pixel we build a list of three scan angles ($\theta,\phi,\psi$) for all samples in that pixel, in other words, the whole scan pattern is transformed to a set of ``Pixel Ordered Detector Angles'' (PODA). Though this step requires significant amount of computation (few  $\times \ 10$\% of total computation), once the satellite pointing information is available, this binning is fixed, so this step need not be repeated.  This means the effective beam for different beam shapes can be recomputed without repeating this step. Also, polarized detectors in Planck are in pairs whose pointing angles are the same and polarization angle differ by a constant threshold, so one need not compute PODA for half of the polarized detectors.

The next step is the load division among computing processes. Though assigning equal number of pixels to each process would lead to the most efficient division of memory, we assign equal numbers of time samples to each process for the following reason: normally the experimental scans are designed in such a way that nearly the whole sky is uniformly sampled, which ensures that even distribution of samples would distribute most of the pixels evenly. However, some pixels (pixels near the ecliptic poles for Planck) can have orders of magnitude more observations than the average. These pixels would slow down the corresponding processes by orders of magnitude and, as a result, the whole computation. Even division of pixels along longitudinal strips would not help, as the distribution of these pixels is highly asymmetric. But in the equal number of samples scheme, these few pixels will be assigned to processes given many fewer pixels than average.  This does not significantly increase the number of pixels in the other processes, since the number of these highly observed pixels is low.

\begin{itemize}

\item {\bf Advantages}: Small I/O overhead, the whole beam is kept loaded in memory, and optionally written to disk only in the end; each process reads and buffers PODA separately and no bottleneck situation occurs; the process is highly scalable, can use thousands of processors; memory management is transparent and possible to keep the whole beam for multiple detectors loaded in memory for the convolutions, without ever writing to the disk.

\item {\bf Disadvantages}: Complicated implementation; requires significant preprocessing and disk space to store PODA.

\end{itemize}

This method is highly efficient and has been implemented for Planck mission. From this point, our effective beam computation refers to pixel space parallelization, unless mentioned otherwise.

\subsection{Implementation for Planck}

We have fully implemented the effective beam method for the Planck mission. The mission specific inputs required for the effective beam computation are the detector characteristics and the pointing information. Planck's detector characteristics---beam parameters, polarization angles, cross polar leakage etc.---are read from a focal plane database. The detector pointing information is extracted from compressed satellite pointing using the Generalized Compressed Pointing~(GCP) library through the MADCAP3 interface~\footnote{\url{http://crd.lbl.gov/~cmc/M3}}.

The beams are stored in binary formats which are accessible to both {\tt C} and {\tt Fortran} standard I/O routines. Each process writes its own binary file, effective beams and other data for each pixel are stored as one ``record'' in one of those files. To preserve compatibility between different computing languages, the record size is kept the same for each pixel, which is determined by maximum $\nlist$. The address of a pixel is stored in a combined index file, which is actually a sky map---each map value encodes the file number and location for that pixel as a single precision integer. A software interface to conveniently load partial or full effective beams has also been developed.

The cost of effective beam computation scales as $\ntod \times \nbeam$, where $\ntod$ is the total number of samples (sum of number of observations at at all pixels). For $15$ months of Planck HFI $143$GHz channel (simulated) data, sampled at $172$Hz (total $6.8 \times 10^9$ samples), with $\npix = 12 \times 2048^2$ pixels, the computation cost is $\sim 130$ CPU hours per detector. For Planck LFI $30$GHz channel, the cost time is $\sim 30$ CPU hours. These costs could be reduced by $\sim 10$\% by using the little less accurate formulae for beam computation provided in section~\ref{subsec:TonlyEB}.

The total size of the effective beam can be estimated as follows: for each pixel with $\nlist$ beam pixels, total volume of data is essentially the number of neighboring pixels plus number of beam matrix elements. So for polarized beams, if we use single precision numbers, the size of each record is approximately $4 (1+6) \nlist = 28\nlist$ Bytes, so the total size of the polarized effective beam will be $28\nlist \npix$ Bytes. For Planck HFI $143$GHz with $\nside=2048$ and $\nlist \sim 225$ the total volume of effective beam turns out to be $\sim 290$GB. This is approximately the disk space required to store the beams, the total amount of (distributed) memory requirement is greater than this, but of the same order. Note that, if we chose to use different beams for unpolarized and polarized components, the polarized effective beam matrices will not be symmetric, and $50$\% more storage would be required. The other factor which may affect the size is the cut-off radius.  Increasing this may improve the accuracy slightly for polarized detectors, but the total beam object size scales as the square of the radius, so the increase in this data volume can be significant.

\begin{figure}
\centering
\includegraphics[width=0.475\textwidth]{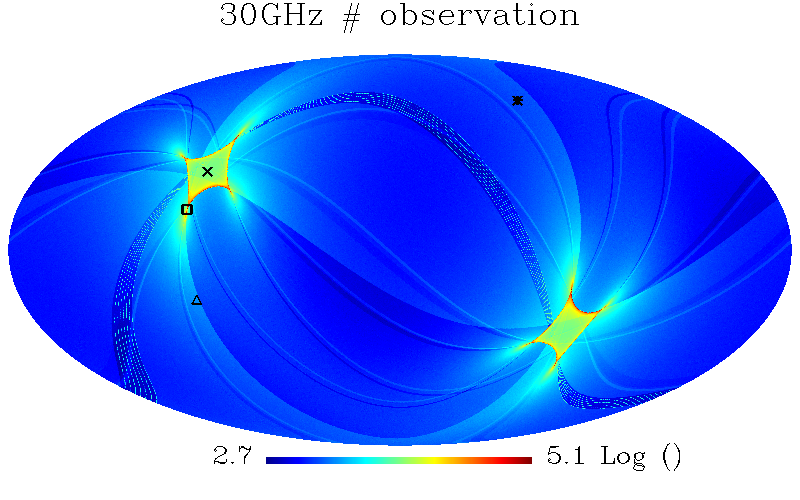}
\caption{Log plot of typical number of observations at each pixel of {\em simulated} Planck sky maps for $\sim 15$ months' scanning. The one shown in the figure is the total number of observations for four $30$GHz detectors in Galactic coordinates. The marked points correspond to the locations of beam and PSF plots shown below.}
\label{fig:nhits}
\end{figure}

We illustrate the features of effective beams and PSFs derived from them by showing a few figures for Planck $30$GHz (simulated data). The distribution in the number of observations due to the scan pattern is shown in Figure~\ref{fig:nhits}.
In Figure~\ref{fig:effBeam}, we show logarithmic plots of effective beams (left) and PSFs (right) at different locations on the sky, where the locations have been marked in Figure~\ref{fig:nhits} (in Galactic coordinates) respectively by ``cross'' (ecliptic North pole),``square'' (a cusp near ecliptic North pole),``triangle'' (a mid-declination point) and ``asterix'' (ecliptic equatorial point). We overlay the isocontours at $50\%$, $10\%$, $1\%$ and $0.1\%$ levels for the numerical maps and also elliptical Gaussian fits to the beams and PSFs. The fitted values, FWHM, ellipticity, orientation angle with respect to the local meridian (vertical axis) and the normalized RMS fit error have been mentioned in the subtitle. The center of the central pixel (effective beam/PSF center) and the center of the fitted Gaussian have been marked with ``$+$'' and ``$\times$'' respectively, when they coincide the markers form an ``eight arm asterix''. These figures tell us that the effective beam / PSF has significant variations across the sky, implying that simplifying assumptions for asymmetric beam studies may not be very accurate. Specially, near the ecliptic poles and the cusps of the scans, the asymmetry can be large.

\begin{figure}
\centering
\includegraphics[width=0.225\textwidth]{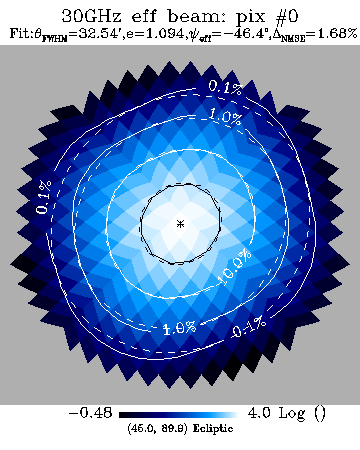}
\includegraphics[width=0.225\textwidth]{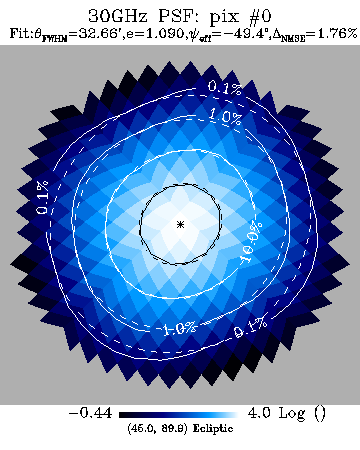}
\includegraphics[width=0.225\textwidth]{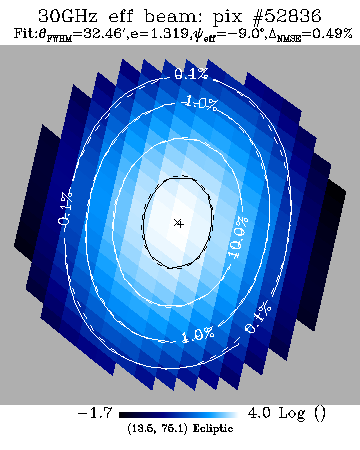}
\includegraphics[width=0.225\textwidth]{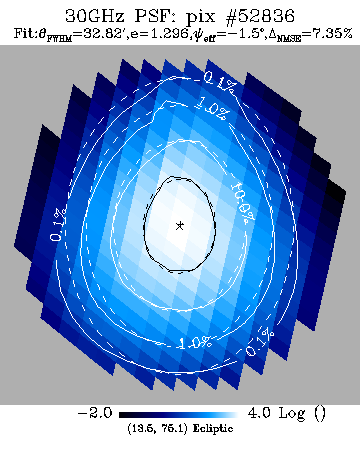}
\includegraphics[width=0.225\textwidth]{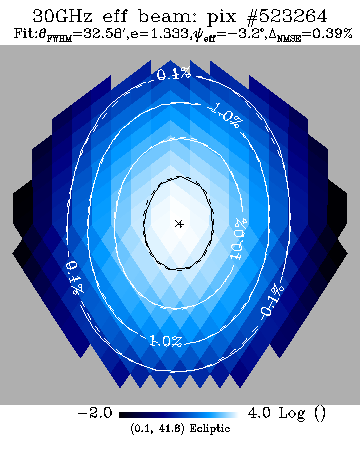}
\includegraphics[width=0.225\textwidth]{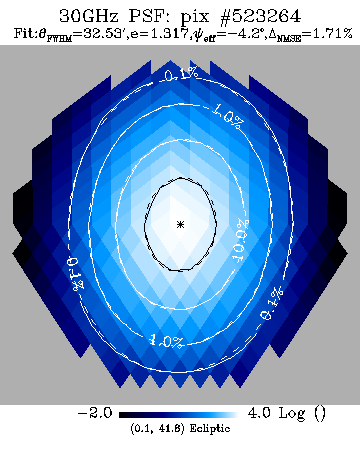}
\includegraphics[width=0.225\textwidth]{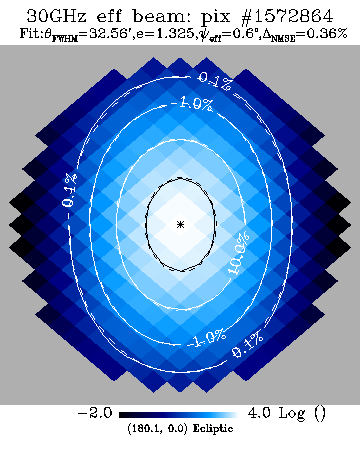}
\includegraphics[width=0.225\textwidth]{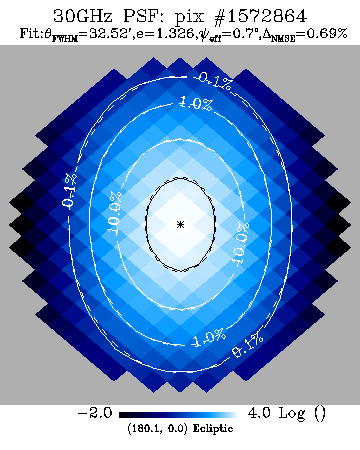}
\caption{Log plots of effective beams (left) and PSFs (right) at four different locations of sky - north ecliptic pole, a cusp near North ecliptic pole, a mid-declination point and ecliptic equator, marked in Figure~\ref{fig:nhits} respectively by ``cross'',``square'',``triangle'' and ``asterix''. The isocontours of the map and elliptical Gaussian fit to the map are overlayed at $50\%$, $10\%$, $1\%$ and $0.1\%$ levels respectively. The center of the effective beam/PSF and that of the Gaussian fit have been marked respectively by ``$+$'' and ``$\times$''.}
\label{fig:effBeam}
\end{figure}

\section{Application of Effective Beams - I: Convolution}
\label{sec:conv}

The main purpose of the effective beams is very fast convolution and the {\tt FEBeCoP} implementation handles it efficiently. Below first we provide some approximate computation cost estimates for the effective beam convolutions; next we test the accuracy by comparing with the results derived using existing Planck simulation software ``level-S'' and analyzed by HEALPix utility {\tt anafast}  \citep{HEALPix}.

\subsection{Efficiency}

Convolution with effective beams not only scales as $\npix \times \nbeam$, the proportionality constant is also very small---of order $1$ for temperature only beam and of order $10$ for polarized beams. So the convolution is very fast, but it is still important to ensure that the implementation is efficient and free of I/O bottlenecks. Among other optimizations, we pre-invert the polarization ``hits'' matrices and ensure that the beams are read in contiguous chunks in order to minimize file access. Our current implementation is now solely limited by map I/O---this is indeed satisfactory that the core computation is faster than the time taken to read a map from disk, but, on the other hand, unfortunate, that we can not make it faster! However, studies that involve online simulation and analysis of maps, e.g., transfer function estimation, where maps need not be read from or written to disk, run almost at the native speed, though they involve some amount of network I/O. We provide some typical computation costs in table~\ref{tab:computationCost} for different beam sizes, assuming that the usual number of HEALPix pixels were used to represent a map.

\begin{table*}[ht]
\centering
{\small
\begin{tabular}{|c|c|c|c|c|c|c|c|c|c|}\hline
& &  \multicolumn{3}{c|}{Temperature only} & \multicolumn{3}{c|}{Polarized}\\\cline{3-8}
Beam & & Effective & \multicolumn{2}{c|}{Computation time} & Effective & \multicolumn{2}{c|}{Computation time} \\\cline{4-5}\cline{7-8}
FWHM & HEALPix & beam size & beam/detector & convolution & beam size & beam/detector & convolution\\
(arcmin) &  $\nside$ & (GB) & (CPU hr) & (CPU min) & (GB) & (CPU hr) & (CPU min)\\\hline\hline
32 & 512 & 7 & 27 & 1.33 & 24 & 31 & 4.4\\ \hline
13 & 1024 & 18 & 42 & 5 & 64 & 48 & 17\\ \hline
7 & 2048 &  90 & 128 & 21& 293 & 132 & 64\\ \hline
\end{tabular}}
\caption{{\em Approximate} computational resource estimates for {\tt FEBeCoP} to compute effective beams for $\sim 15$ months' of observation (with cut-off radius of $2 \, \times$ FWHM) and to convolve using the (precomputed) effective beams on the NERSC computing facility.}
\label{tab:computationCost}
\end{table*}

\subsection{Accuracy: Comparison with Existing Planck Simulations}

Next, we assess the accuracy of the convolution by comparing with the simulated maps generated using the standard Planck Level-S TOD simulation pipeline in combination with the Springtide or MADAM map maker~\citep{mapMakerComp}. Starting from the focal plane database, the satellite scan strategy and (the spherical harmonic transforms of) an input map, we generate Total Convolution Data Cube with $m_{\rm max} = 2$ for symmetric beams and $m_{\rm max} = 14$ for asymmetric beams. The TCDC are then interpolated by {\tt multimod}~\citep{levelS}, to generate TOD at each scan sample. These (noiseless) TODs are then binned by a mapmaker to generate maps. The effective beams are computed using the same pointing and focal plane database and convolved with (the pixelized version of) the input map. The comparison between the maps and the power spectra for two Planck channels, $30$GHz (all detectors: $27$a, $27$b, $28$a, $28$b) and $143$GHz (four detectors: $1$a, $1$b, $4$a, $4$b), are shown below. Among all the Planck beams, the $30$GHz ones have the highest asymmetry (ellipticity $\sim 1.35$ - $1.40$) and the $143$GHz ones have the lowest asymmetry (ellipticity  $\sim 1.02$ - $1.11$).

\begin{figure}
\centering
\includegraphics[width=0.225\textwidth]{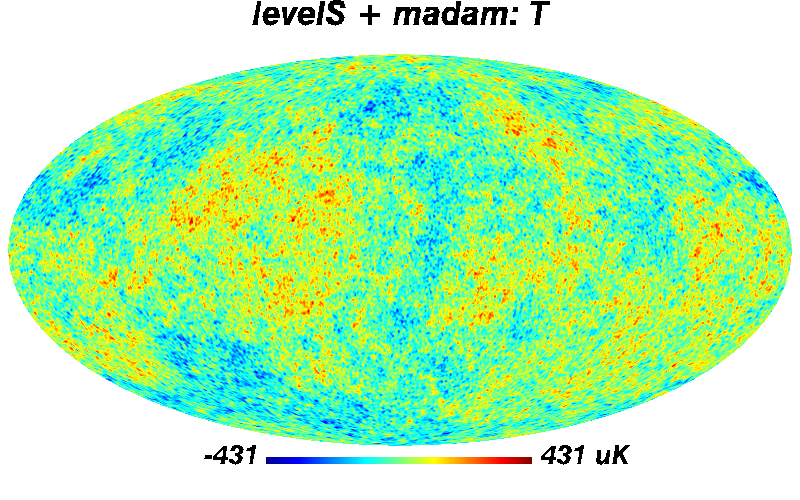}
\includegraphics[width=0.225\textwidth]{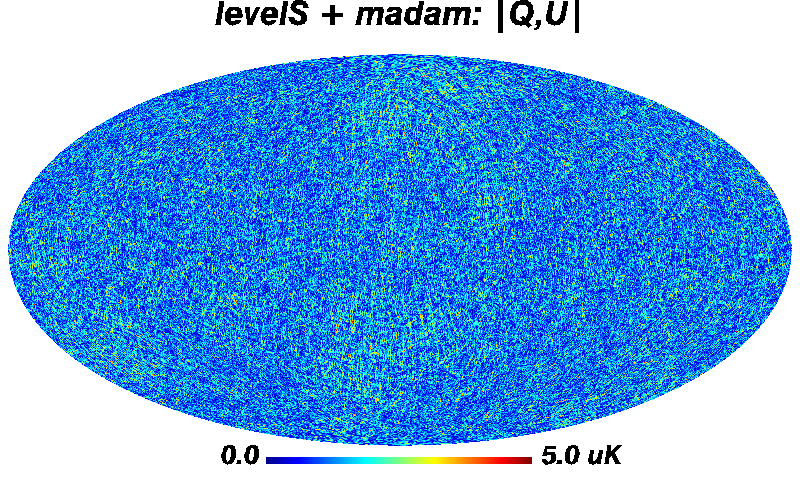}\\
\includegraphics[width=0.225\textwidth]{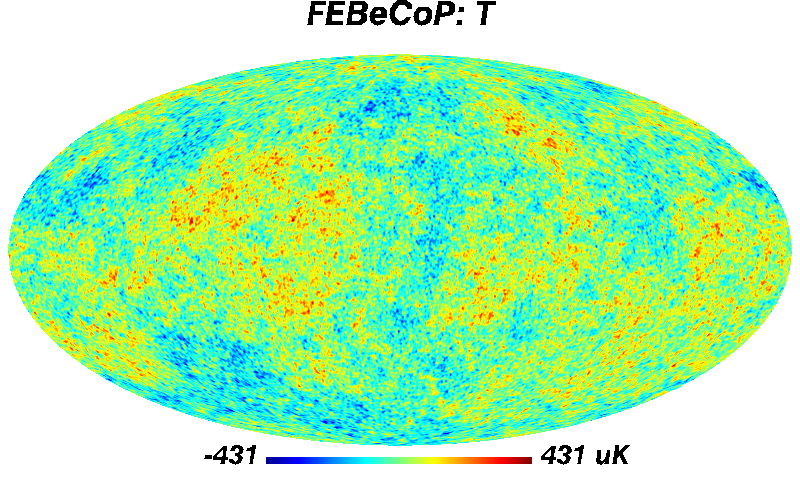}
\includegraphics[width=0.225\textwidth]{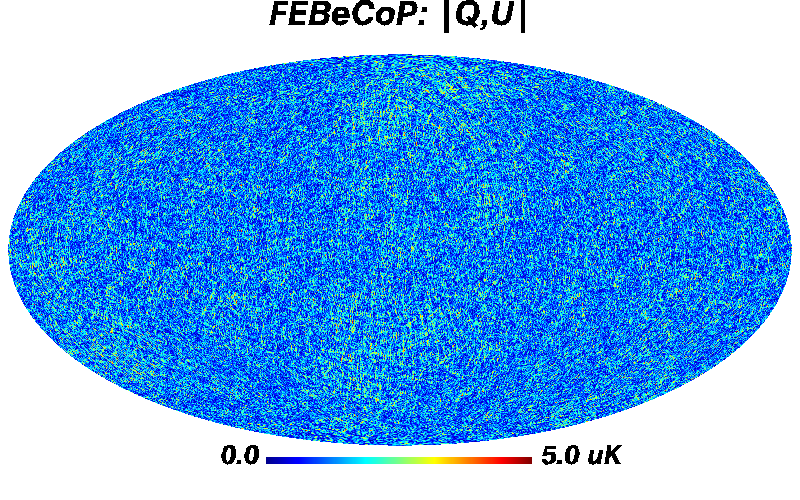}\\
\includegraphics[width=0.225\textwidth]{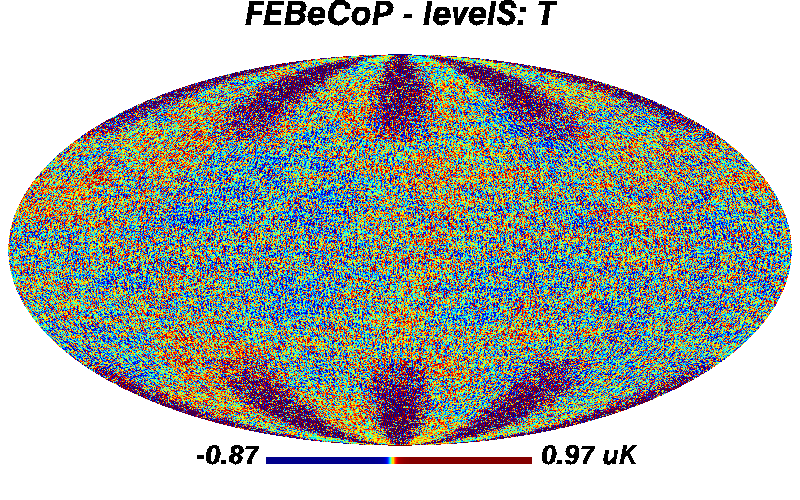}
\includegraphics[width=0.225\textwidth]{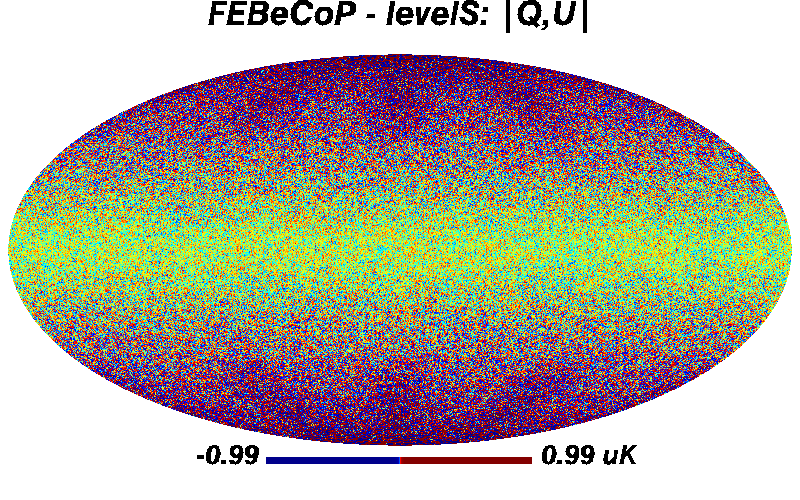}
\caption{Comparison between simulated temperature (left panel) and polarization (right panel) anisotropy maps for Planck 30GHz channels obtained using Level-S (top) and {\tt FEBeCop} (middle). The maps are visually quite similar, the difference map (bottom) are also shown using a histogram equalized color map (to amplify the contrast in the image). The fractional RMS difference for temperature is $0.04$\% and for polarization is $0.9$\%.}
\label{fig:compFullSky}
\end{figure}

\begin{figure*}[h]
\centering
\includegraphics[width=0.225\textwidth]{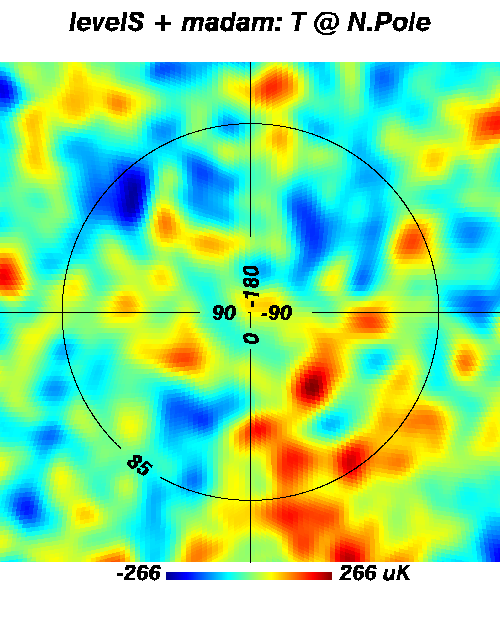}
\includegraphics[width=0.225\textwidth]{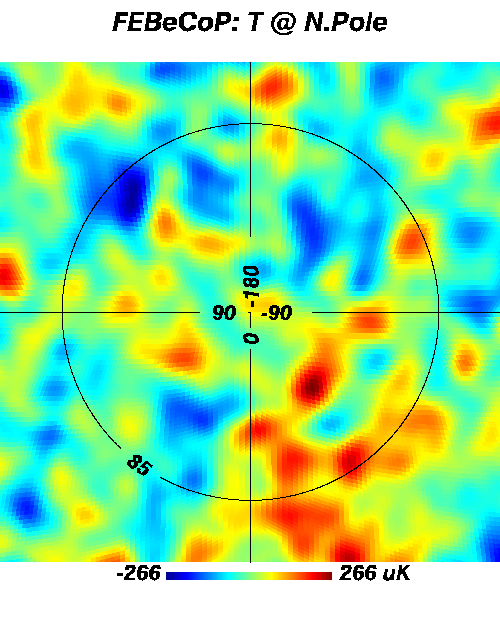}
\includegraphics[width=0.225\textwidth]{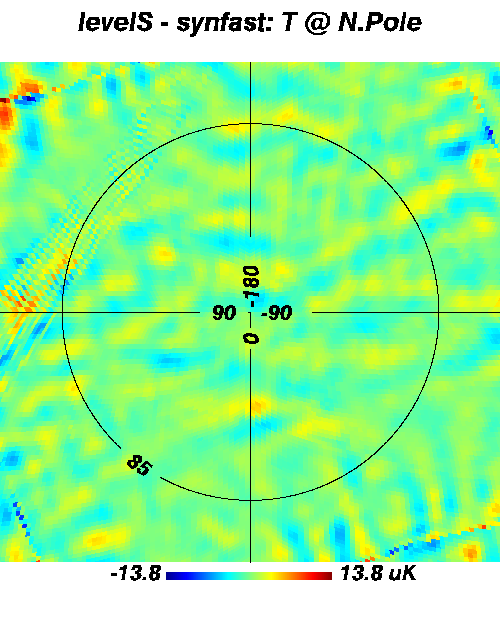}
\includegraphics[width=0.225\textwidth]{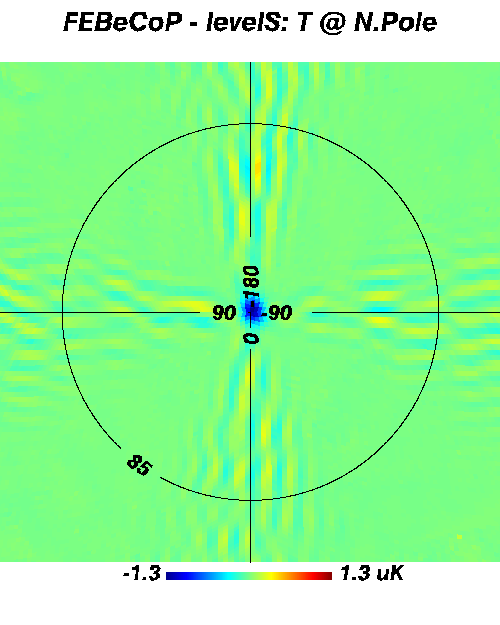}
\includegraphics[width=0.225\textwidth]{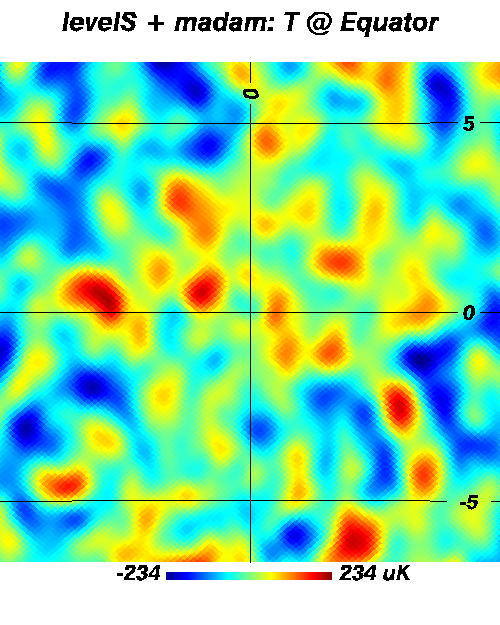}
\includegraphics[width=0.225\textwidth]{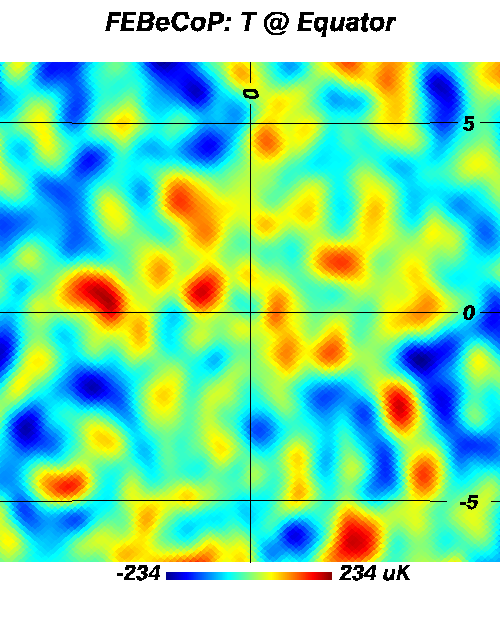}
\includegraphics[width=0.225\textwidth]{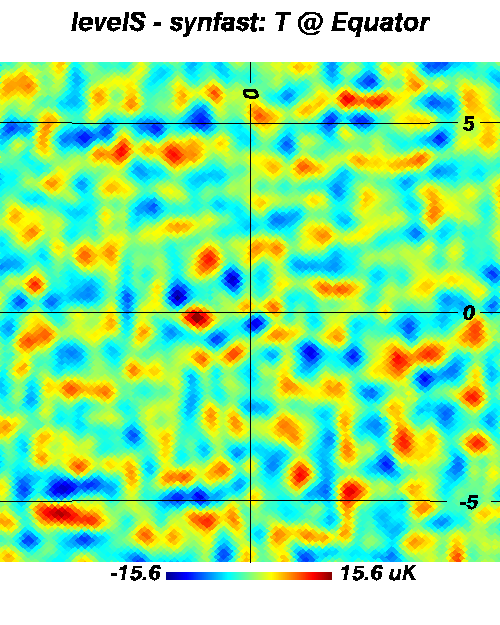}
\includegraphics[width=0.225\textwidth]{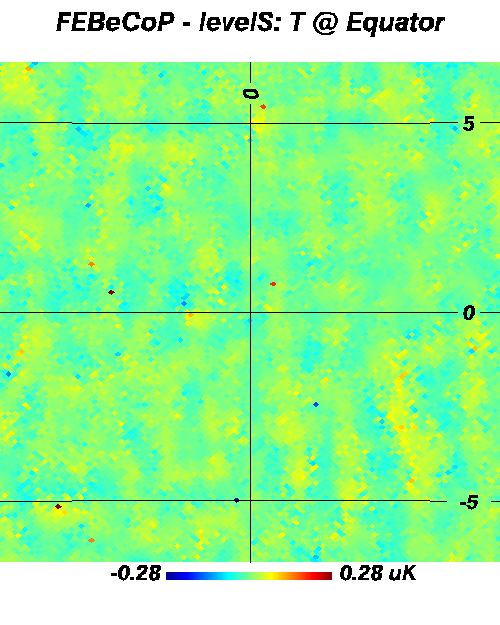}
\includegraphics[width=0.225\textwidth]{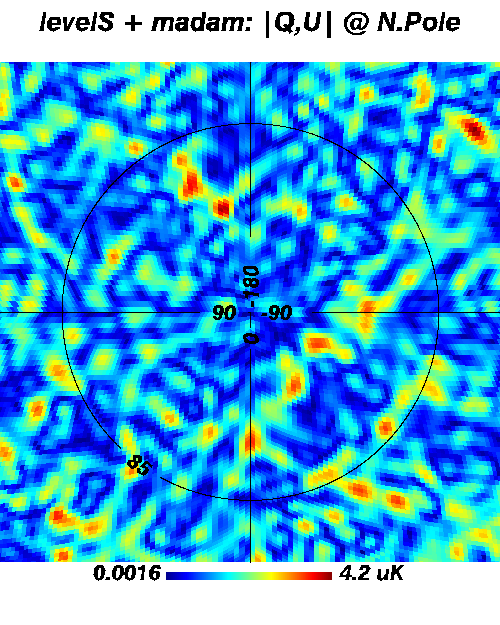}
\includegraphics[width=0.225\textwidth]{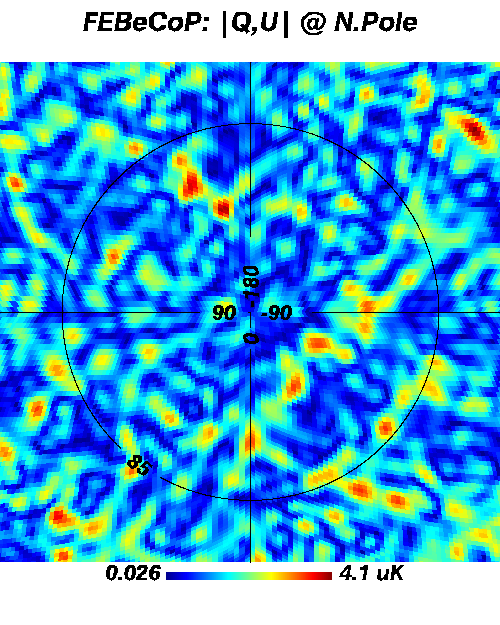}
\includegraphics[width=0.225\textwidth]{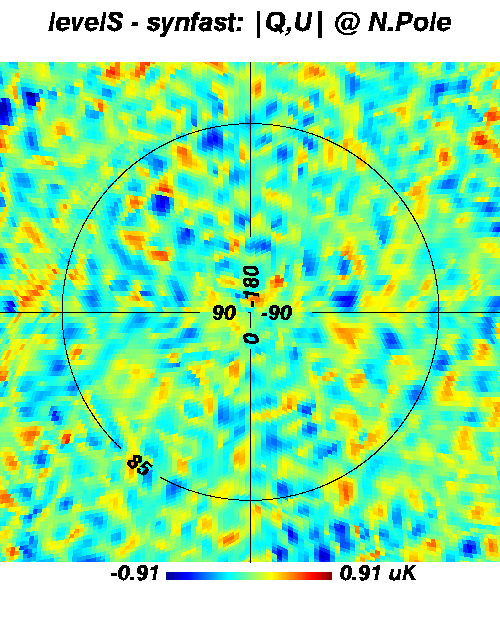}
\includegraphics[width=0.225\textwidth]{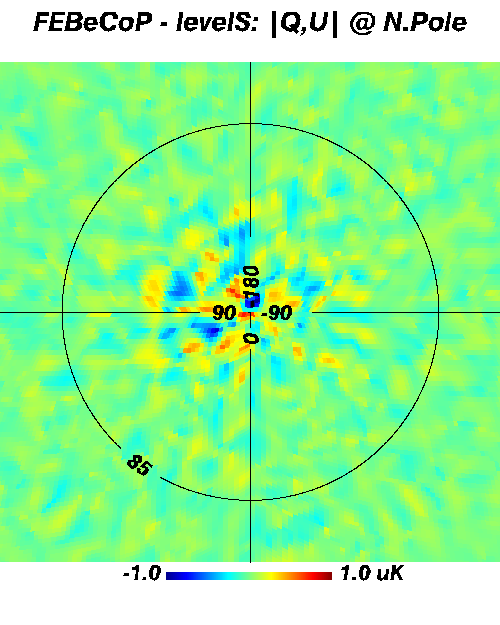}
\includegraphics[width=0.225\textwidth]{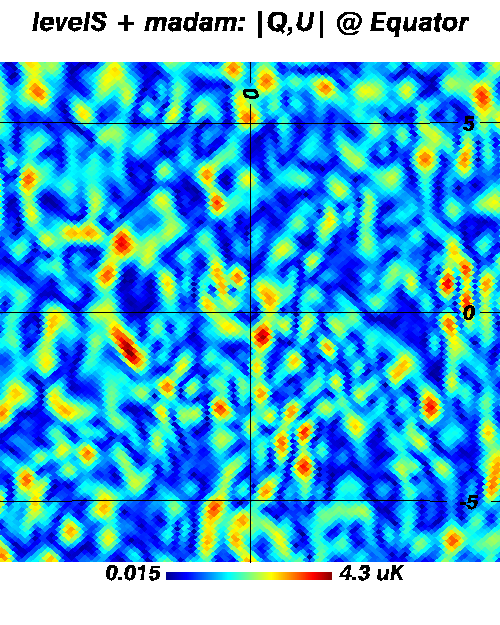}
\includegraphics[width=0.225\textwidth]{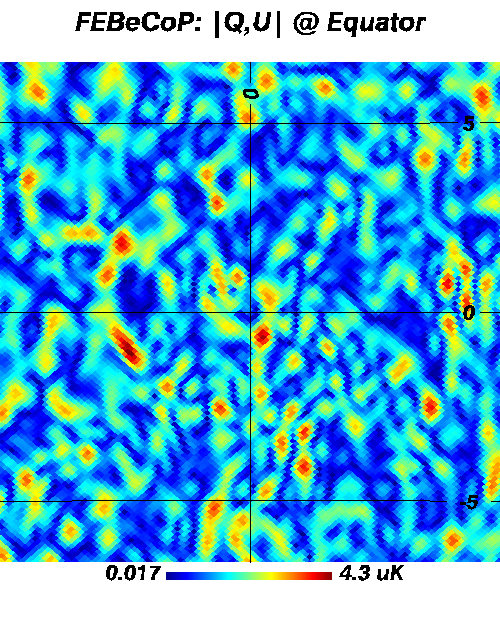}
\includegraphics[width=0.225\textwidth]{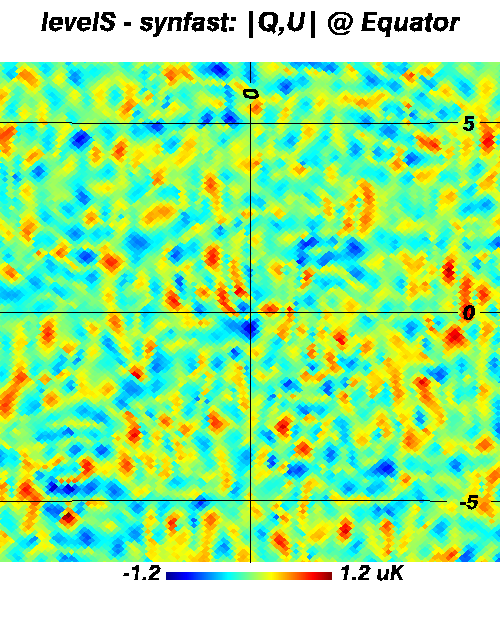}
\includegraphics[width=0.225\textwidth]{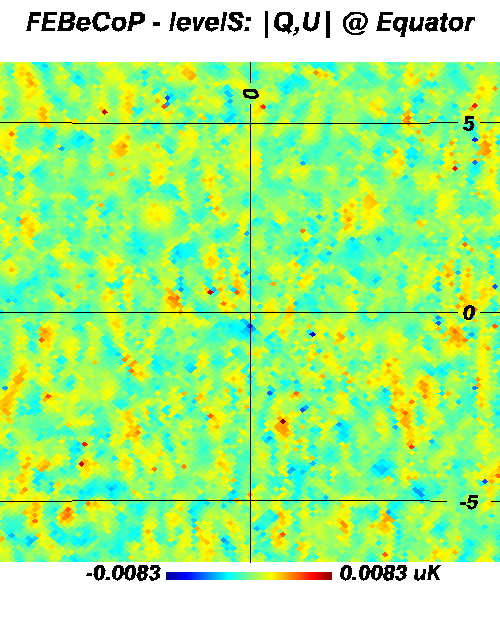}
\caption{Zoomed version of comparison of maps from Level-S (column \# 1) and {\tt FEBeCoP} (column \# 2) at two different locations---the ecliptic North pole ( row \# 1 \& 3) and equator (row \# 2 \& 4). The difference between the Level-S maps and the {\tt synfast} generated symmetric Gaussian smoothed maps (column \# 3) show that the effect of asymmetric beams is significant. Finally the difference between Level-S and {\tt FEBeCoP} is very small (column \# 4), showing that {\tt FEBeCoP} simulates the realistic effects as accurate as the detailed Level-S simulations.}
\label{fig:diffMap}
\end{figure*}

\begin{figure}
\centering
\includegraphics[width=0.45\textwidth]{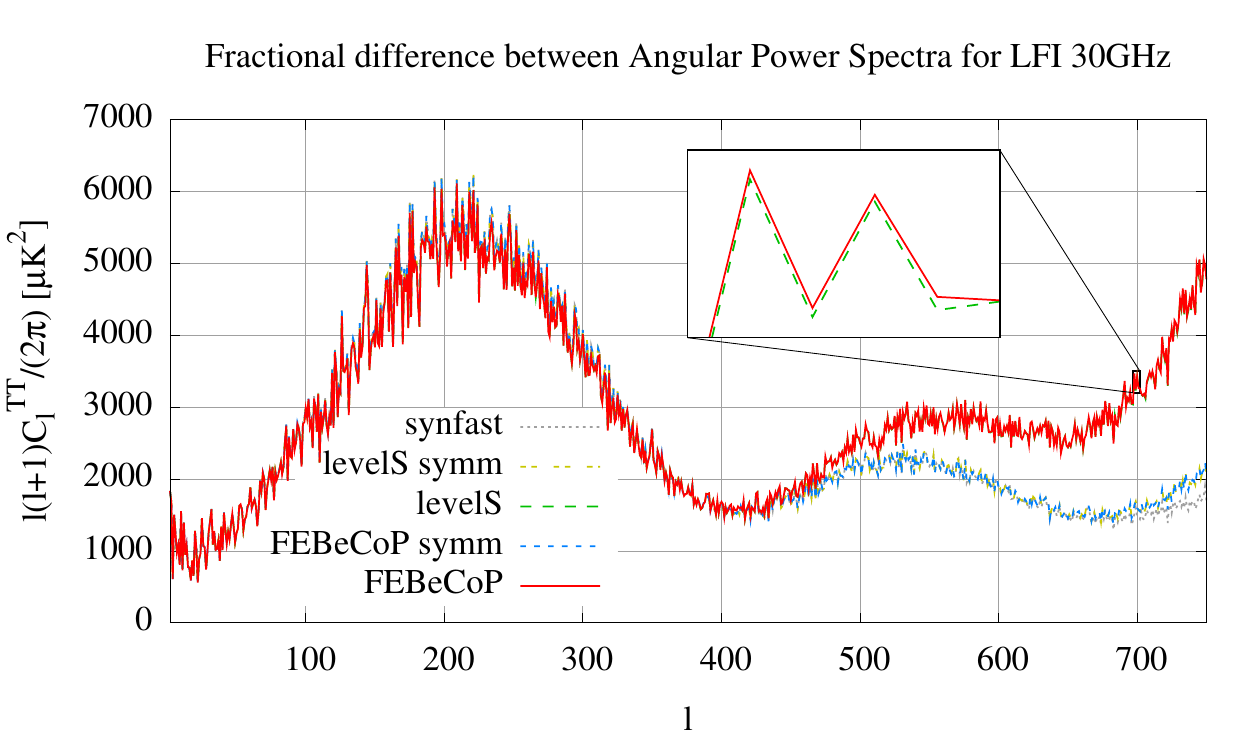}
\includegraphics[width=0.45\textwidth]{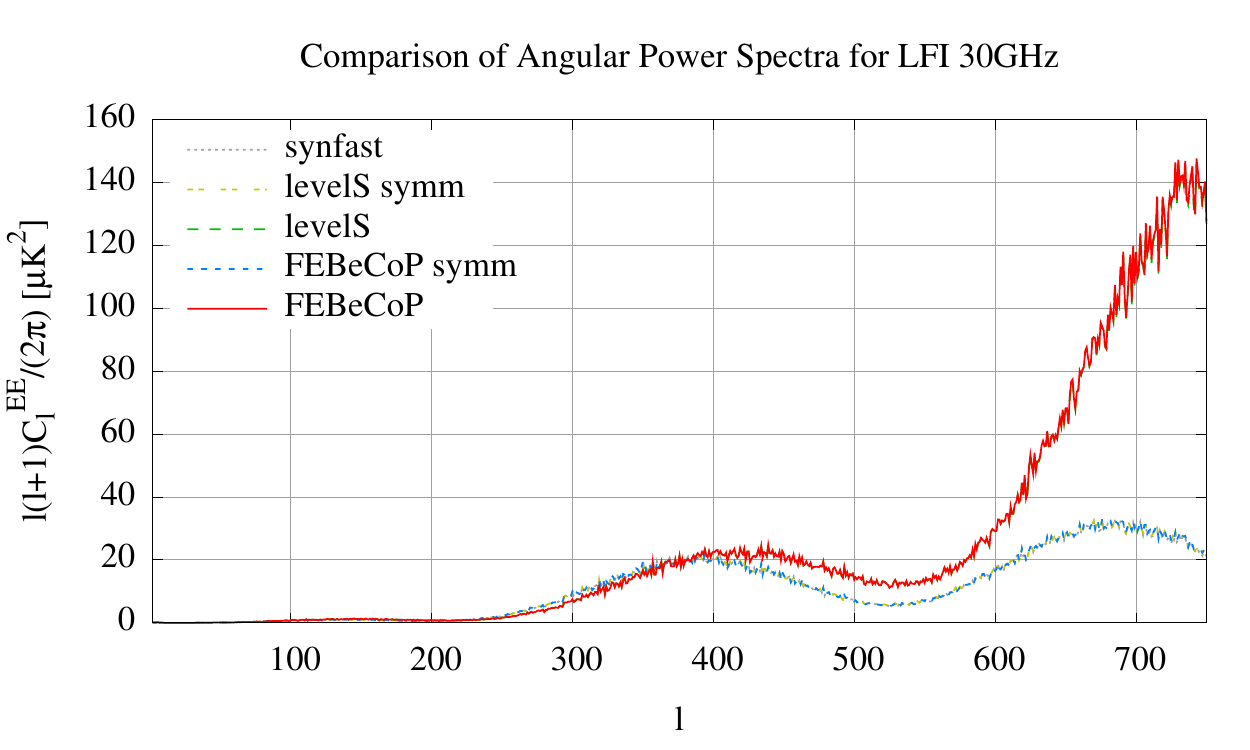}
\includegraphics[width=0.45\textwidth]{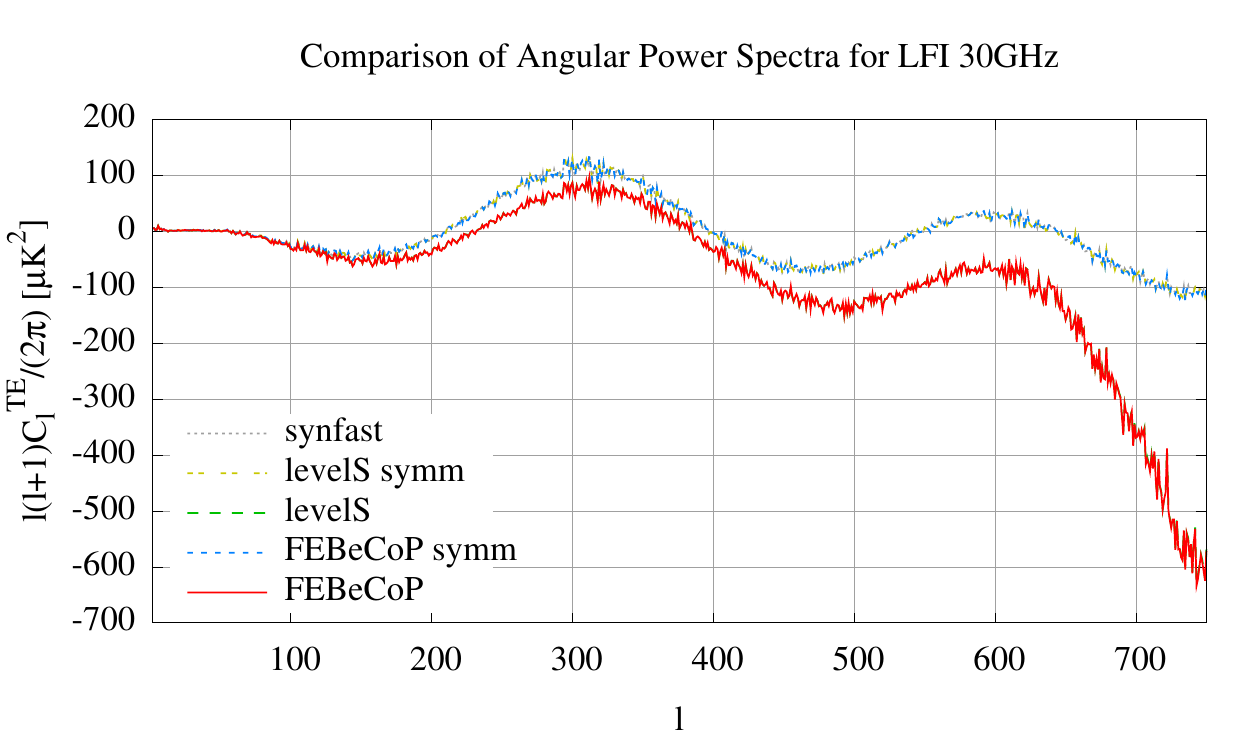}
\caption{Comparison of $TT$ (top), $EE$ (middle) and $TE$ (bottom) power spectra for LFI $30$GHz: the main purpose of this plot is to show that Level-S and {\tt FEBeCoP} spectra match very well for both both symmetric and asymmetric beams (the lines are on top of each other). This plot shows that if the detector intrinsic beam is perfectly symmetric, the spectra are close to those of {\tt synfast} generated symmetric Gaussian smoothed maps.}
\label{fig:30Cl}
\end{figure}

\begin{figure}
\centering
\includegraphics[width=0.45\textwidth]{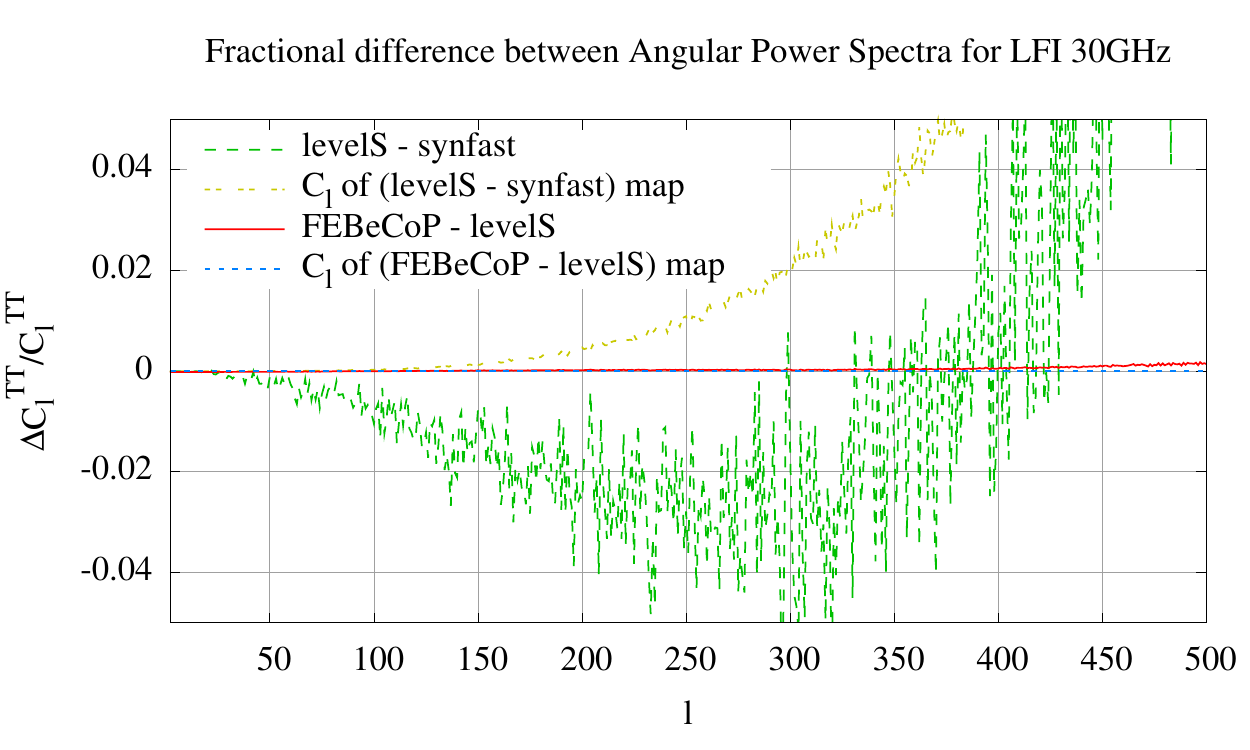}
\includegraphics[width=0.45\textwidth]{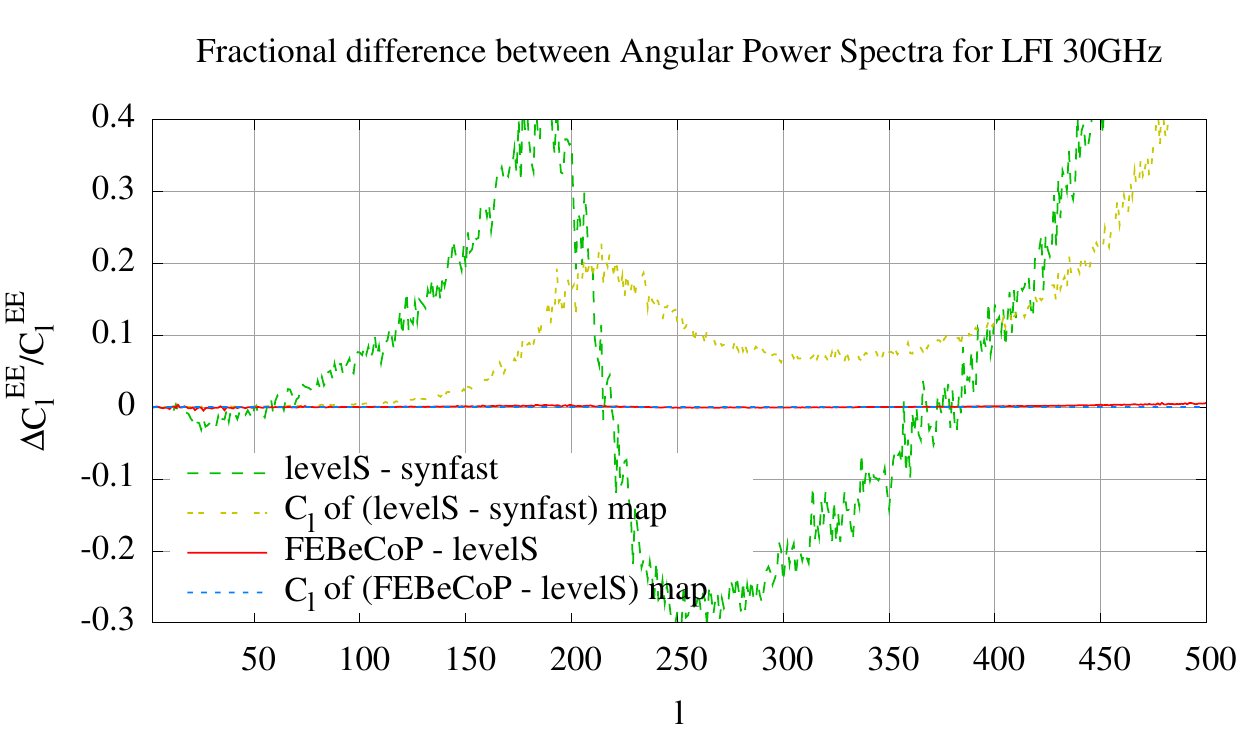}
\caption{Difference between power spectra of the maps and power spectra of difference maps for LFI $30$GHz.  While the difference between Level-S maps and ideal symmetric beam convolved map using {\tt synfast} is large (dashed), few $10$\%, the difference level between {\tt FEBeCoP} and Total Convolution + Level-S + MADAM is less than $\sim 0.5$\% at almost all multipoles (solid). This shows that the effect of asymmetric beam for polarization is large and {\tt FEBeCoP} can imitate the effect accurately.}
\label{fig:diff30Cl}
\end{figure}

\begin{figure}
\centering
\includegraphics[width=0.45\textwidth]{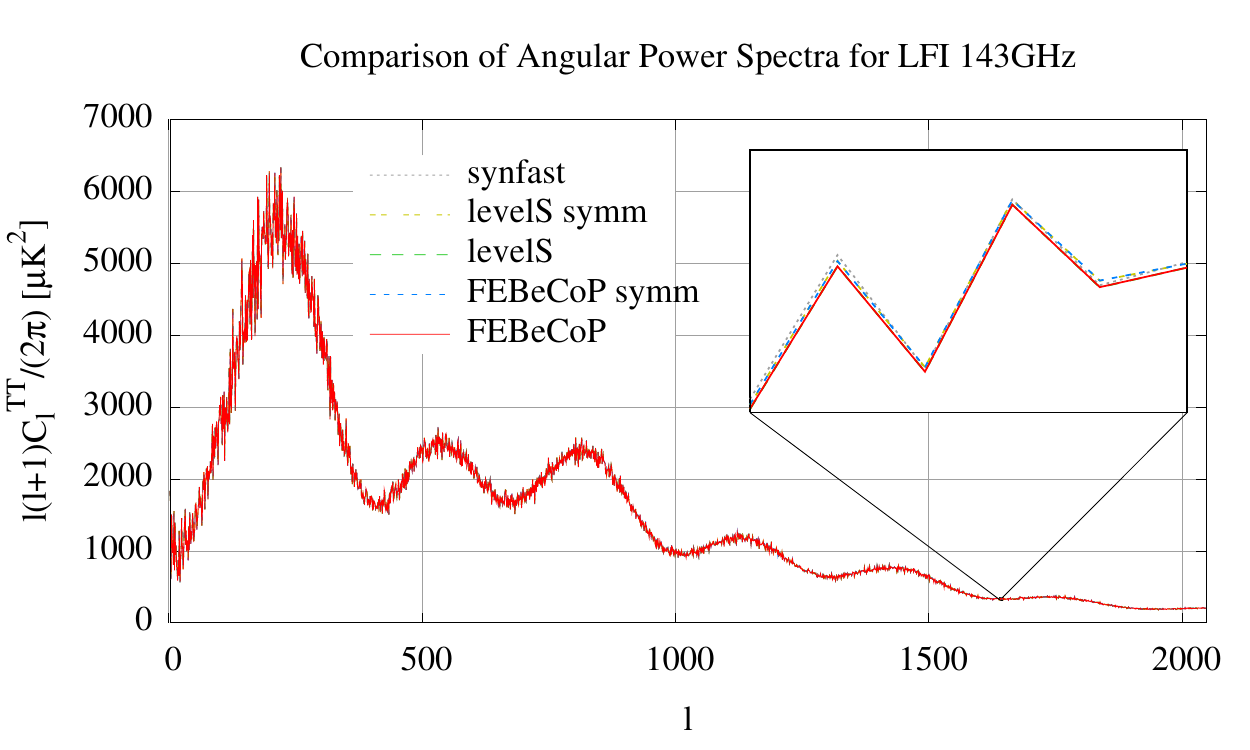}
\includegraphics[width=0.45\textwidth]{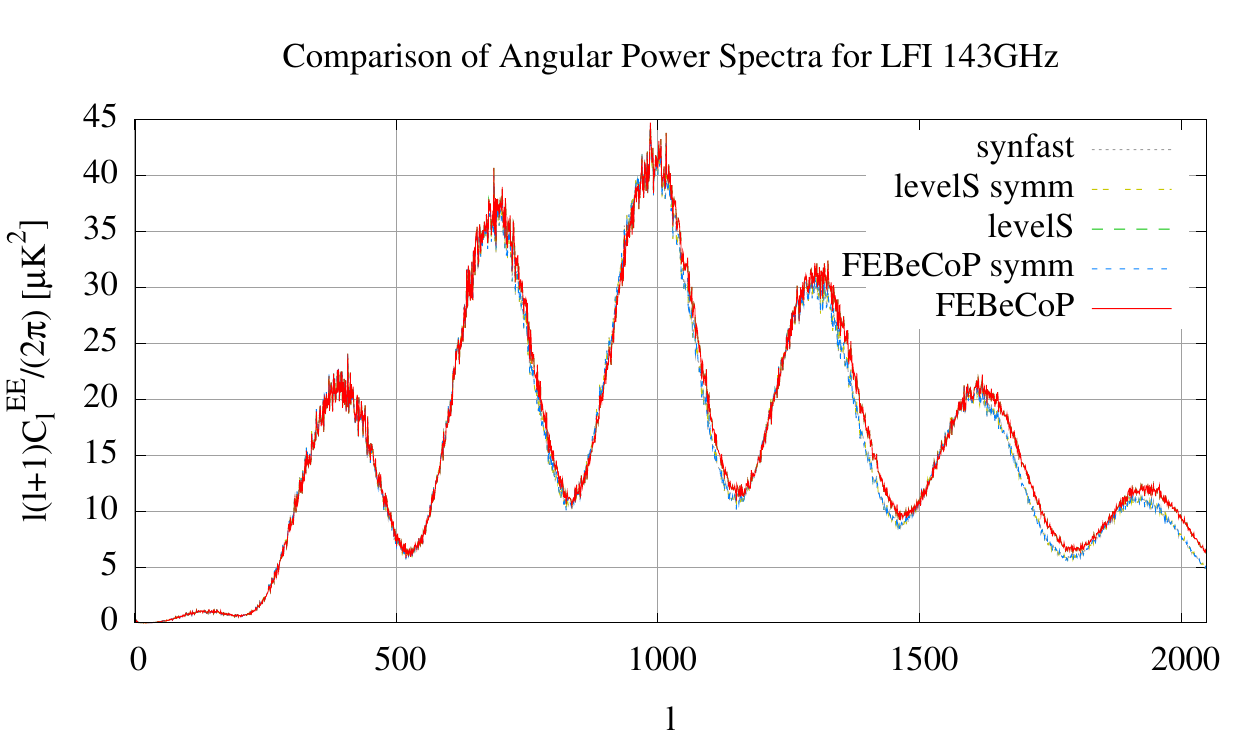}
\includegraphics[width=0.45\textwidth]{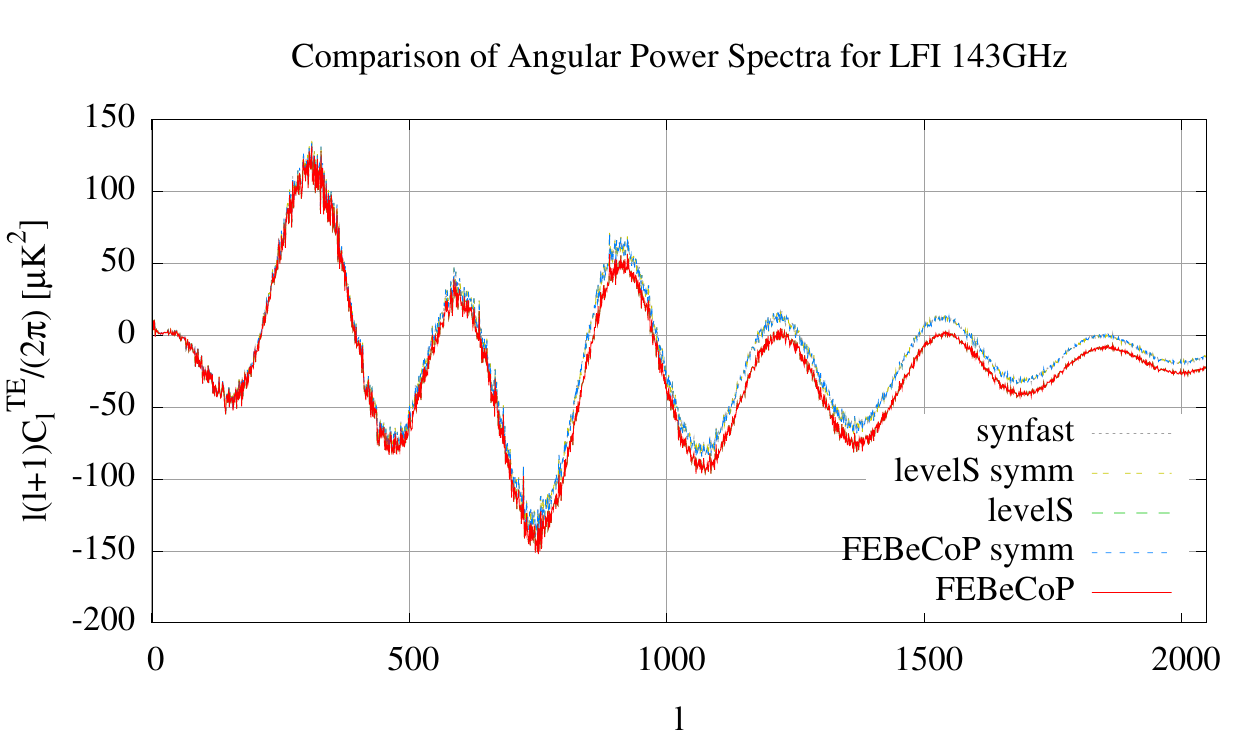}
\caption{Comparison of $TT$ (top), $EE$ (middle) and $TE$ (bottom) power spectra for HFI $143$GHz: the main purpose of this plot is to show that Level-S and {\tt FEBeCoP} spectra for both symmetric and asymmetric match very well (the lines are on top of each other). This plot also shows that if the detector intrinsic beam is perfectly symmetric, the spectra are close to those of {\tt synfast} generated symmetric Gaussian smoothed maps.}
\label{fig:143Cl}
\end{figure}

\begin{figure}
\centering
\includegraphics[width=0.45\textwidth]{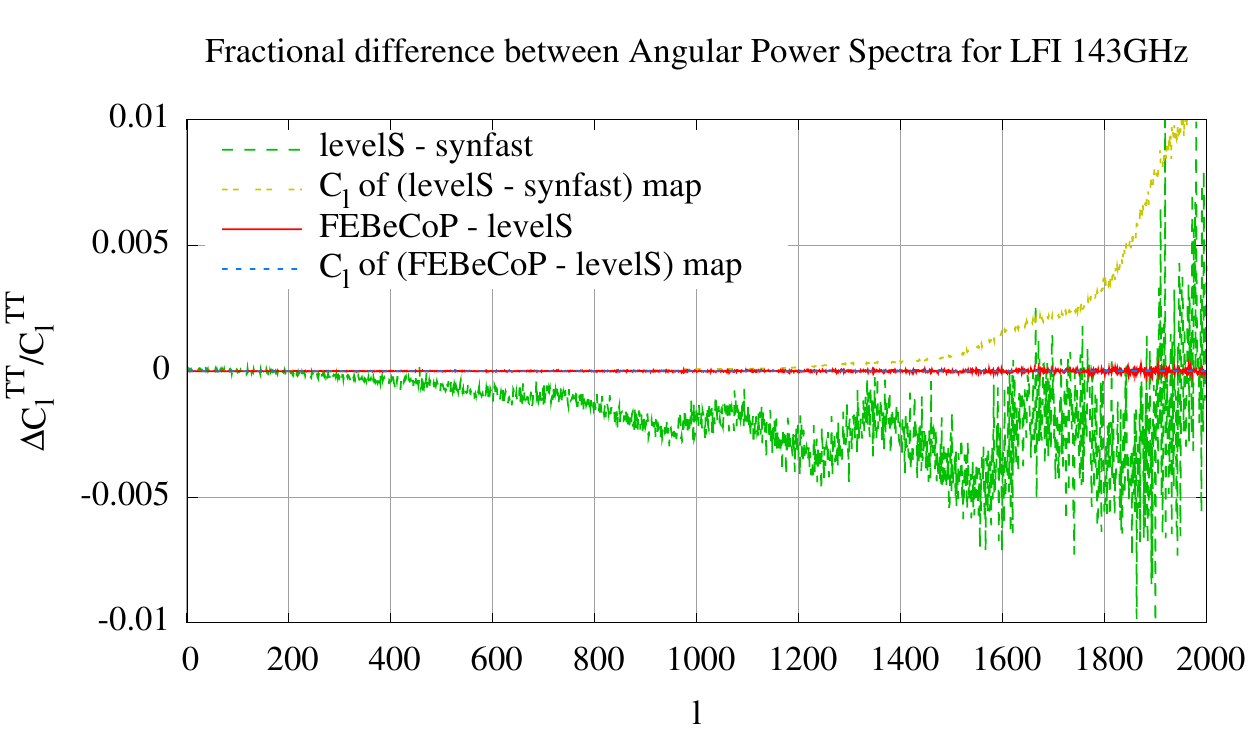}
\includegraphics[width=0.45\textwidth]{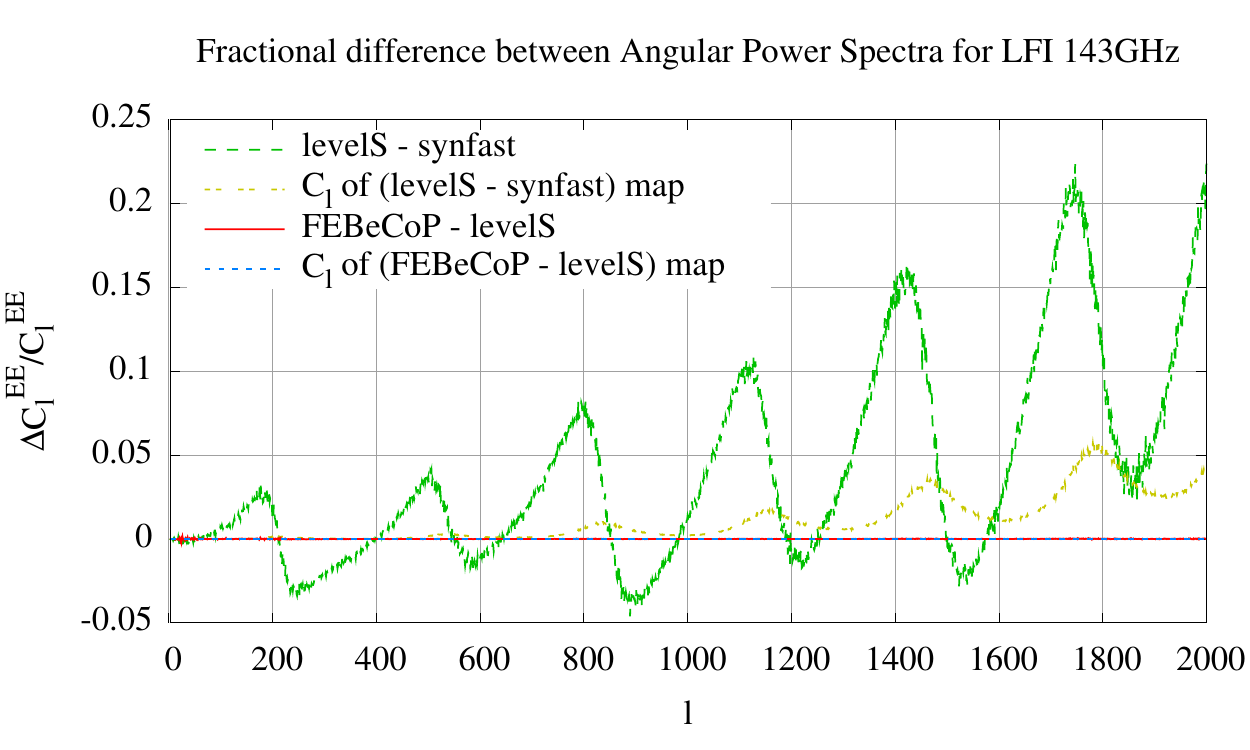}
\caption{Difference between power spectra of the maps and power spectra of difference maps for HFI $143$GHz.  While the difference between Level-S maps and ideal symmetric beam convolved map using {\tt synfast} is large (dashed), as much as $20$\%, the difference level between {\tt FEBeCoP} and Total Convolution + Level-S + MADAM is less than $\sim 0.02$\% at almost all multipoles (solid). This shows that the effect of asymmetric beam for polarization is large and {\tt FEBeCoP} can imitate the effect accurately.}
\label{fig:diff143Cl}
\end{figure}

We first show simulated full sky temperature and polarization maps for Planck $30$GHz channel using Level-S and {\tt FEBeCoP} and their differences in left and right panels of Figure~\ref{fig:compFullSky}. The maps from the two methods are clearly similar and the difference is very small---the fractional RMS difference for temperature maps is few $\times \ 0.01\%$ and for polarization amplitude the difference is few $\times \ 0.1\%$ (note that the contrast in the difference maps have been amplified by using a histogram equalized color map).
We have also plotted zoomed versions of the above maps for North polar and equatorial regions in Figure~\ref{fig:diffMap}. The agreement between FEBeCoP and levelS is good in all the places, expect for only few pixels (among millions) right at the poles, which is not significant. (We omit these plots for 143 GHz, where they are uninformative.) The power spectra of the maps from two methods for both $30$GHz and $143$GHz\footnote{We observed that for the $143$GHz channel, increasing the cut-off radius from $2$ times FWHM to $2.25$ times FWHM, increases the polarization accuracy by a small amount. However, the storage also increases by $25$\%, so there is some small trade-off involved in the choice of parameters.} channels are shown in [Figure~\ref{fig:30Cl} and \ref{fig:143Cl}], the spectra are almost indistinguishable (the lines nearly coincide).  Their fractional differences reveal more details [Figure~\ref{fig:diff30Cl} and \ref{fig:diff143Cl}]. What is important to notice in these comparisons is that the difference between the effective beam convolved maps and the Level-S plus mapmaker maps, and their power spectra, is much smaller than the corresponding difference between a Level-S and symmetric Gaussian smoothed map, obtained using the HEALPix utility {\tt synfast}  \citep{HEALPix}. Especially for polarization the symmetric beam assumption systematic is large ($\sim {\rm few} \ 10\%$), while the effective beam error is $\lesssim 0.1\%$. Even for symmetric detector intrinsic beams, small asymmetry is introduced to the effective beams by scan and that too is faithfully reproduced by {\tt FEBeCoP}. This can be seen in the top panel of Figure~\ref{fig:30Cl}. Here the spectrum of theoretical symmetric Gaussian convolved map computed using {\tt synfast} differs from the spectrum of levelS map with symmetric beam and simulated Planck scan, but the spectrum of FEBeCoP map for the same inputs lies exactly on top of the LevelS spectrum. These exercises demonstrate that effective beam convolution provides very good accuracy in mimicking the systematic effects in the detectors.



\section{Application of Effective Beams - II: To Study Collective Effects of Beams}
\label{sec:effBl}

Effective beams are not merely useful for fast convolution, they can provide a more realistic way of understanding and interpreting systematic effects. To demonstrate the potential applications, we provide two examples in this section.

\subsection{Statistics of the Effective Beams}

\begin{figure}
\centering
\includegraphics[width=0.45\textwidth]{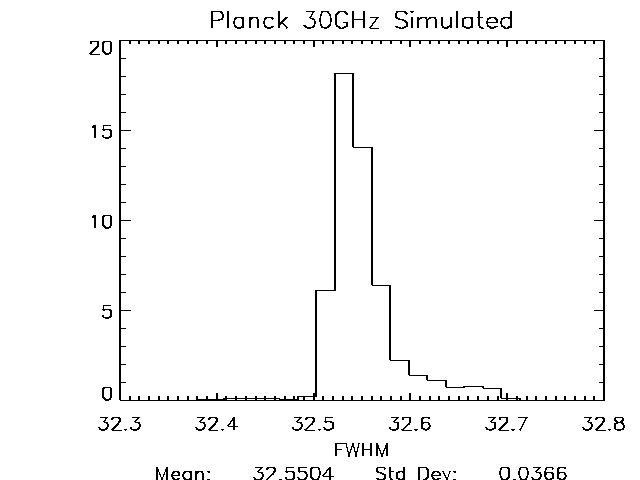}
\includegraphics[width=0.45\textwidth]{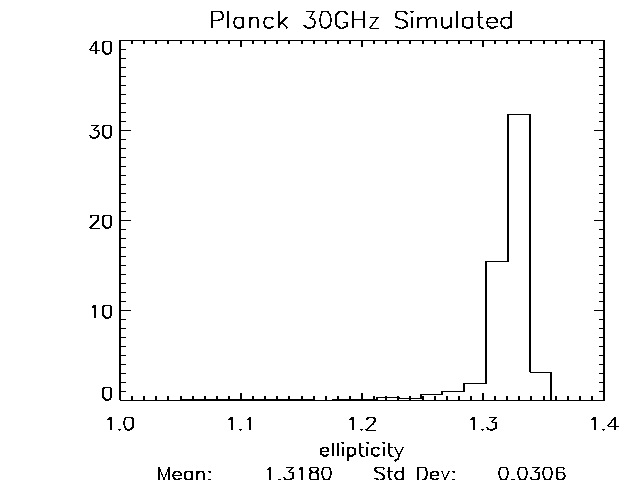}
\includegraphics[width=0.45\textwidth]{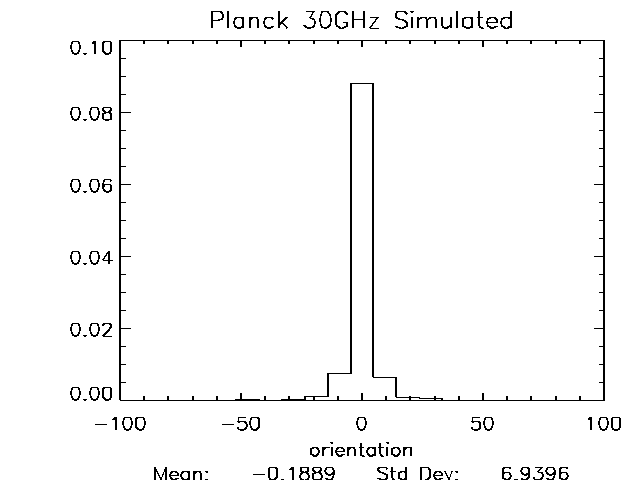}
\caption{Variation of elliptical Gaussian fit parameters to $30$GHz simulated effective beams is shown in this figure. FWHM is in arcmin and the orientation angle is in degrees. The average orientation angle is close to zero in this plot due to the overall detector geometry specific to this case. Tools developed for {\tt FEBeCoP} allows fast and easy applications to study the statistical properties of effective beams.}
\label{fig:effBeamStat}
\end{figure}

Complex combination of scanning, beam and pixel shapes lead to different shapes of effective beams and PSFs, few examples were shown in Figure~\ref{fig:effBeam}. Software tools developed for {\tt FEBeCoP} readily allow one to obtain useful statistics of the beams and PSFs. In Figure~\ref{fig:effBeamStat} we show histograms characterizing the variation of elliptical Gaussian fit parameters to the $30$GHz simulated effective beams, sampled at HEALPix pixel centers of $\nside=16$ (this method ensures uniform sampling, though it has a very high probability of missing the ``cusp'' regions in the scan pattern, where the beams/PSFs can be drastically different).
Typically the effective beams are few percent bigger than the intrinsic detector beams as their centers are spread over a pixel, adding extra smoothing.

\subsection{Understanding Observed Power Spectra}
In the previous section we presented comparisons between symmetric and asymmetric beam convolved maps. The observed power spectra of these maps were different by few percent. There is, however, no trivial way of interpreting this difference---the reason is that the effective beams have varying shapes across the sky. Previous attempts to explain this systematic deviation using mathematical models assume constant beam shape (with possibly rotation) across the sky, though the real scan strategy does not have a simple analytical model. This problem can be partially or completely tackled by studying the effective beams. Here we use a leading order approximation to show that the ratio of temperature power spectra observed using asymmetric and symmetric beams can be modeled using the rotated spherical harmonic transforms of the effective beams.

For a symmetric beam of spatially non-varying size, it is trivial to show that the observed power spectrum $\cobs_l$ is related to the true power spectra through the relation
\begin{equation}
\cobs_l \ = \ B_l^2 \, C_l \, ,
\end{equation}
where $B_l$ are the Legendre transforms of the beam. In general, though, the effective beam shapes are not constant and they vary across the sky, as shown in Figure~\ref{fig:effBeam}. So the transfer function is no longer as simple as $B_l^2$. The easiest way to estimate the transfer function is by performing Monte Carlo simulations and estimate the ratio $\cobs_l/C_l$ for each multipole, which is, in fact, one of the motivations for the effective beam approach. However, to extract more insights directly form the effective beams, we crudely estimate average $B_l^2$ of effective beams by taking some sample effective beams from different latitude, computing the rotated harmonic transform and incorporating the leading order correction following the algebra in \citet{mitra04}.

Even though these approximations are crude, the results look quite satisfactory.
\begin{figure}
\centering
\includegraphics[width=0.45\textwidth]{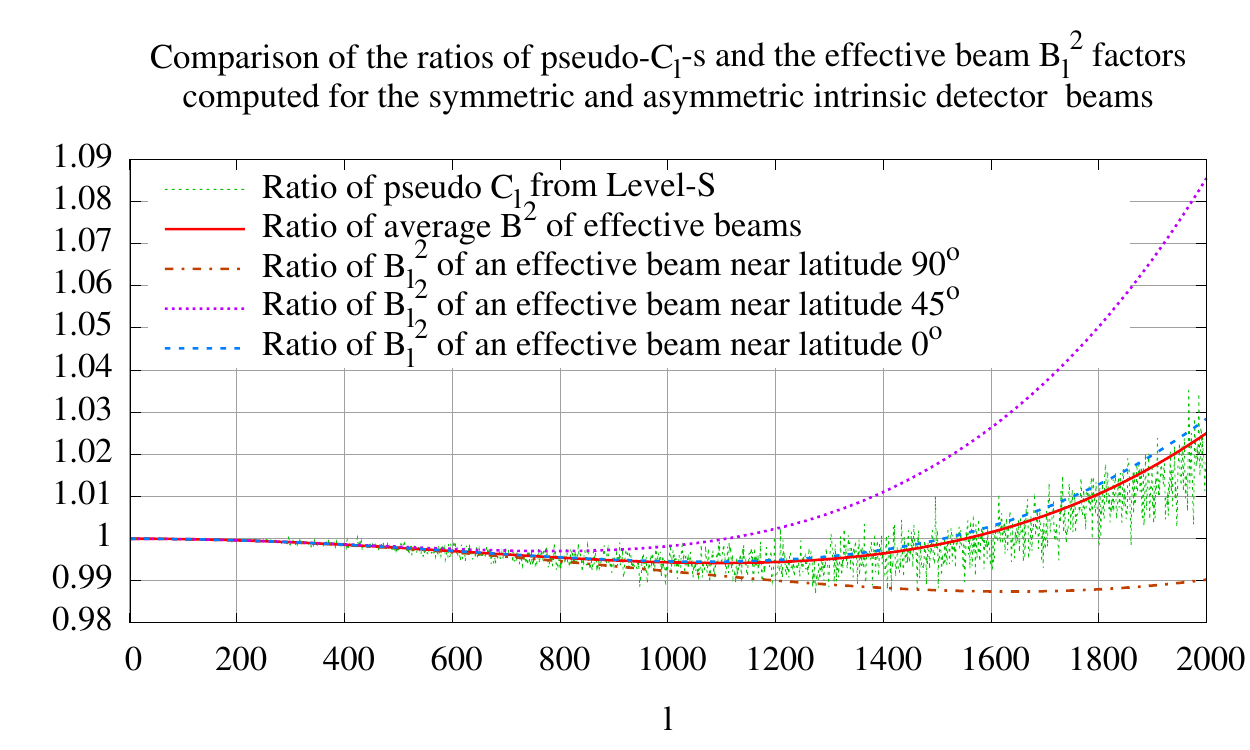}
\caption{Illustration of approximate modeling of the transfer function using harmonic transforms of the effective beams. Dotted: the ratio of observed power spectra of simulated maps for Planck 143GHz channel \#5 for asymmetric and symmetric detector intrinsic beams. Solid: Ratio of average effective $B_l^2$ obtained from the effective beams for the same simulations, which reasonably matches the $C_l$ ratio. The $B_l^2$  ratio at certain latitudes (no average) are also shown to highlight the variation of the effective beams across the sky.}
\label{fig:effBlSqRatio}
\end{figure}
In Figure~\ref{fig:effBlSqRatio}, we plotted the ratio of power spectra of simulated maps for Planck 143-5 channel with asymmetric and symmetric beams. For the same two cases, the ratio of $B_l^2$ estimated from effective beams is overlaid on the plot. Clearly the match is quite good for demonstration purposes. There are accuracy issues at high multipoles due to many reasons---for example, the formula we used is valid for beams with reflection symmetry, though the effective beams do not have that symmetry---the effective beams are constructed from elliptic or circular beams which have that symmetry, but since the beam centers and orientations are not distributed uniformly in a pixel, the reflection symmetry is slightly broken, which may introduce extra first order correction terms.

The ratio of $B_l^2$ of effective beams at certain latitudes (not averaged) are also overlaid to show that the characteristics of the effective beams vary significantly at different sky locations, which may break simplified assumptions often used in simulations. The variation of $B_l^2$ with latitude need not represent any trend, as the effective beams can vary significantly from pixels to pixels and also planck scanning is not completely azimuthally symmetric.

\section{Conclusions}
\label{sec:concl}

Realistic simulations of high precision experiments play an extremely important role in  studying systematic effects of the instrument. There is no exception for scanning CMB experiments, like Planck. In the last several years there has been significant success in developing an end-to-end (E2E) simulation pipeline to simulate very realistic Planck time-order-data (TOD) going through the full scan strategy. However, such simulations involve a very  detailed computational procedure, and are usually computationally prohibitive in applications to adequate Monte Carlo studies 
(that is repetitive many thousand runs, or more). Many CMB analyses do not require full TOD for the study of systematic effects, as simulated realistic observed maps provide all the ingredients required for these studies. In fact, these studies may require many more simulations than it is feasible to obtain by performing the E2E simulations, and then going through mapmaking. Beam asymmetry is one of  the most important potential instrumental systematic effect in Planck. Pseudo-$C_l$, and other power spectrum estimators could make use of large number of realistic sky map simulations to estimate the correction needed to estimate the true power spectra. This was the motivation for developing a pixel based convolution method using effective beams, which we have implemented using the computer code {\tt FEBeCoP}. The method precomputes the effective beams and makes it possible to perform very fast convolutions to simulate realistic sky maps. {\tt FEBeCoP} also provides point spread functions (PSFs) at every pixel on the sky, which is otherwise non-trivial to obtain. We have successfully constrained the computation cost to a moderate value as well as managed to properly utilize the available distributed computing power. The computation requirement for {\tt FEBeCoP} is modest as compared to the modern scientific computing facilities available for Planck analyses. Due to the simplicity of {\tt FEBeCoP} interface and its efficiency, we believe that it would be very easy and useful to incorporate effective beam convolution in many systematic effect studies.

\section{Acknowledgments}

We would like to thank B.~Crill, M.~Seiffert and F.~Bouchet for useful discussions on Planck beams. We acknowledge C.~Cantalupo and T.~Kisner  for support  with MADcap3 interface, I.~O'Dwyer  for assistance in devising data formats for storage of  effective beams, and  R.~Keskitalo for  aid in usage of MADAM mapmaker. Part of the research described in this paper was carried out at the Jet Propulsion Laboratory, California Institute of Technology, under a contract with the National Aeronautics and Space Administration.


\bibliographystyle{apj}
\bibliography{EffBeam}

\begin{thebibliography}{35}
\expandafter\ifx\csname natexlab\endcsname\relax\def\natexlab#1{#1}\fi

\bibitem[{{A.~D. Miller {\it et~al.}}(1999)}]{mill99}
{A.~D. Miller {\it et~al.}} 1999, Astrophys. J., 524, L1

\bibitem[{{Armitage-Caplan} \& {Wandelt}(2009)}]{PReBeaM}
{Armitage-Caplan}, C., \& {Wandelt}, B.~D. 2009, ApJs, 181, 533

\bibitem[{{Ashdown} {et~al.}(2007){Ashdown}, {Baccigalupi}, {Balbi},
  {Bartlett}, {Borrill}, {Cantalupo}, {de Gasperis}, {G{\'o}rski}, {Hivon},
  {Keih{\"a}nen}, {Kurki-Suonio}, {Lawrence}, {Natoli}, {Poutanen}, {Prunet},
  {Reinecke}, {Stompor}, {Wandelt}, \& {The Planck CTP Working
  Group}}]{mapMakerComp}
{Ashdown}, M.~A.~J., {et~al.} 2007, \aap, 467, 761

\bibitem[{{Ashdown} {et~al.}(2009){Ashdown}, {Baccigalupi}, {Bartlett},
  {Borrill}, {Cantalupo}, {de Gasperis}, {de Troia}, {G{\'o}rski}, {Hivon},
  {Huffenberger}, {Keih{\"a}nen}, {Keskitalo}, {Kisner}, {Kurki-Suonio},
  {Lawrence}, {Natoli}, {Poutanen}, {Pr{\'e}zeau}, {Reinecke}, {Rocha},
  {Sandri}, {Stompor}, {Villa}, {Wandelt}, \& {The Planck Ctp Working
  Group}}]{Trieste08}
---. 2009, \aap, 493, 753

\bibitem[{Burigana \& Saez(2003)}]{burigana03}
Burigana, C., \& Saez, D. 2003, Astron. \& Astrophys., 409, 423

\bibitem[{{C.~L. Kuo {\it et~al.}}(2004)}]{acbar}
{C.~L. Kuo {\it et~al.}} 2004, Astrophys. J., 600, 32

\bibitem[{{Challinor} {et~al.}(2000){Challinor}, {Fosalba}, {Mortlock},
  {Ashdown}, {Wandelt}, \& {G{\'o}rski}}]{ChallinorEtAl-00}
{Challinor}, A., {Fosalba}, P., {Mortlock}, D., {Ashdown}, M., {Wandelt}, B.,
  \& {G{\'o}rski}, K. 2000, \prd, 62, 123002

\bibitem[{{Dupac} \& {Tauber}(2005)}]{dupac05}
{Dupac}, X., \& {Tauber}, J. 2005, A\&A, 430, 363

\bibitem[{{E. Torbet {\it et~al.}}(1999)}]{toco}
{E. Torbet {\it et~al.}} 1999, Astrophys. J., 521, L79

\bibitem[{{Fosalba} {et~al.}(2002){Fosalba}, {Dor{\'e}}, \&
  {Bouchet}}]{Fosalba01}
{Fosalba}, P., {Dor{\'e}}, O., \& {Bouchet}, F.~R. 2002, \prd, 65, 063003

\bibitem[{{G{\'o}rski} {et~al.}(2005){G{\'o}rski}, {Hivon}, {Banday},
  {Wandelt}, {Hansen}, {Reinecke}, \& {Bartelmann}}]{HEALPix}
{G{\'o}rski}, K.~M., {Hivon}, E., {Banday}, A.~J., {Wandelt}, B.~D., {Hansen},
  F.~K., {Reinecke}, M., \& {Bartelmann}, M. 2005, ApJ, 622, 759

\bibitem[{{Hinshaw} {et~al.}(2007){Hinshaw}, {Nolta}, {Bennett}, {Bean},
  {Dor{\'e}}, {Greason}, {Halpern}, {Hill}, {Jarosik}, {Kogut}, {Komatsu},
  {Limon}, {Odegard}, {Meyer}, {Page}, {Peiris}, {Spergel}, {Tucker}, {Verde},
  {Weiland}, {Wollack}, \& {Wright}}]{wmap3T}
{Hinshaw}, G., {et~al.} 2007, ApJs, 170, 288

\bibitem[{{Huffenberger} {et~al.}(2010){Huffenberger}, {Crill}, {Lange},
  {G{\'o}rski}, \& {Lawrence}}]{Huffenberger10}
{Huffenberger}, K.~M., {Crill}, B.~P., {Lange}, A.~E., {G{\'o}rski}, K.~M., \&
  {Lawrence}, C.~R. 2010, \aap, 510, A58+

\bibitem[{{Jones} {et~al.}(2007){Jones}, {Montroy}, {Crill}, {Contaldi},
  {Kisner}, {Lange}, {MacTavish}, {Netterfield}, \& {Ruhl}}]{JonesEtAl-06}
{Jones}, W.~C., {et~al.} 2007, \aap, 470, 771

\bibitem[{{Larson} {et~al.}(2010){Larson}, {Dunkley}, {Hinshaw}, {Komatsu},
  {Nolta}, {Bennett}, {Gold}, {Halpern}, {Hill}, {Jarosik}, {Kogut}, {Limon},
  {Meyer}, {Odegard}, {Page}, {Smith}, {Spergel}, {Tucker}, {Weiland},
  {Wollack}, \& {Wright}}]{WMAP}
{Larson}, D., {et~al.} 2010, ArXiv e-prints

\bibitem[{{Maffei} {et~al.}(2010){Maffei}, {Noviello}, {Murphy}, {Ade},
  {Lamarre}, {Bouchet}, {Brossard}, {Catalano}, {Colgan}, {Gispert}, {Gleeson},
  {Jones}, {McAuley}, {Lange}, {Longval}, {Pajot}, {Peacocke},
  {et~al.}}]{Maffei10}
{Maffei}, B., {et~al.} 2010, A\&A

\bibitem[{{Mitra} {et~al.}(2009){Mitra}, {Sengupta}, {Ray}, {Saha}, \&
  {Souradeep}}]{mitra09}
{Mitra}, S., {Sengupta}, A.~S., {Ray}, S., {Saha}, R., \& {Souradeep}, T. 2009,
  Mon. Not. Roy. Astron. Soc., 394, 1419

\bibitem[{{Mitra} {et~al.}(2004){Mitra}, {Sengupta}, \& {Souradeep}}]{mitra04}
{Mitra}, S., {Sengupta}, A.~S., \& {Souradeep}, T. 2004, \prd, 70, 103002

\bibitem[{{P. de~Bernardis {\it et~al.}}(2000)}]{boom}
{P. de~Bernardis {\it et~al.}} 2000, Nature, 404, 955

\bibitem[{{Penzias} \& {Wilson}(1965)}]{PenziasWilson}
{Penzias}, A.~A., \& {Wilson}, R.~W. 1965, ApJ, 142, 419

\bibitem[{{Prezeau} \& {Reinecke}(2010)}]{conviqt}
{Prezeau}, G., \& {Reinecke}, M. 2010, ApJ, accepted for publication

\bibitem[{{Reinecke} {et~al.}(2006){Reinecke}, {Dolag}, {Hell}, {Bartelmann},
  \& {En{\ss}lin}}]{levelS}
{Reinecke}, M., {Dolag}, K., {Hell}, R., {Bartelmann}, M., \& {En{\ss}lin},
  T.~A. 2006, \aap, 445, 373

\bibitem[{{Rocha} {et~al.}(2010){Rocha}, {Pagano}, {G{\'o}rski},
  {Huffenberger}, {Lawrence}, \& {Lange}}]{Rocha09}
{Rocha}, G., {Pagano}, L., {G{\'o}rski}, K.~M., {Huffenberger}, K.~M.,
  {Lawrence}, C.~R., \& {Lange}, A.~E. 2010, \aap, 513, A23+

\bibitem[{{S. Hanany {\it et~al.}}(2000)}]{maxima}
{S. Hanany {\it et~al.}} 2000, Astrophys. J. Lett., 545, L5

\bibitem[{{S. Padin {\it et~al.}}(2001)}]{dasi}
{S. Padin {\it et~al.}} 2001, Astrophys. J. Lett., 549, L1

\bibitem[{{Sandri} {et~al.}(2002{\natexlab{a}}){Sandri}, {Bersanelli}, \&
  {Burigana}}]{sandri:2002}
{Sandri}, M., {Bersanelli}, M., \& {Burigana}, C.~{\it et al.}.
  2002{\natexlab{a}}, in American Institute of Physics Conference Series,
  Experimental Cosmology at Millimetre Wavelengths, ed. M. de Petris \& M.
  Gervasi, 616, 242

\bibitem[{{Sandri} {et~al.}(2002{\natexlab{b}}){Sandri}, {Bersanelli}, \&
  {Burigana}}]{sandri:2009}
---. 2002{\natexlab{b}}, in American Institute of Physics Conference Series,
  Experimental Cosmology at Millimetre Wavelengths, ed. M. de Petris \& M.
  Gervasi, 616, 242

\bibitem[{{Smoot} {et~al.}(1992){Smoot}, {Bennett}, {Kogut}, {Wright}, {Aymon},
  {Boggess}, {Cheng}, {de Amici}, {Gulkis}, {Hauser}, {Hinshaw}, {Jackson},
  {Janssen}, {Kaita}, {Kelsall}, {Keegstra}, {Lineweaver}, {Loewenstein},
  {Lubin}, {Mather}, {Meyer}, {Moseley}, {Murdock}, {Rokke}, {Silverberg},
  {Tenorio}, {Weiss}, \& {Wilkinson}}]{COBEDMR}
{Smoot}, G.~F., {et~al.} 1992, ApJL, 396, L1

\bibitem[{Souradeep \& Ratra(2001)}]{TR2001}
Souradeep, T., \& Ratra, B. 2001, Astrophys. J., 560, 28

\bibitem[{{T.~J. Pearson {\it et~al.}}(2003)}]{cbi}
{T.~J. Pearson {\it et~al.}} 2003, Astrophys. J., 591, 556

\bibitem[{{Tauber} {et~al.}(2009){Tauber}, {Mandolesi}, {Puget}, \&
  {Bersanelli}}]{planckmission09}
{Tauber}, J., {Mandolesi}, N., {Puget}, J.~L., \& {Bersanelli}, M. et~al, P.
  2009, A\&A

\bibitem[{{Tauber} {et~al.}(2010)}]{PlanckOptics}
{Tauber}, J.~A., {et~al.} 2010, \aap, 520, A2+

\bibitem[{{The Planck Collaboration}(2006)}]{bluebook06}
{The Planck Collaboration}. 2006, {The Scientific Programme of Planck}

\bibitem[{{Wandelt} \& {G{\'o}rski}(2001)}]{WandeltGorski-00}
{Wandelt}, B.~D., \& {G{\'o}rski}, K.~M. 2001, \prd, 63, 123002

\bibitem[{{Yurchenko} {et~al.}(2004){Yurchenko}, {Murphy}, \&
  {Lamarre}}]{yurchenko:2004}
{Yurchenko}, V.~B., {Murphy}, J.~A., \& {Lamarre}, J.~M. 2004, Proceedings of
  the SPIE, ed. J. C. Mather, 5487, 542

\end{thebibliography}

\end{document}